\documentclass[useAMS, usenatbib]{mn2e}
\usepackage{epsfig}
\usepackage{graphicx}

\newcommand{\degrees}[1]{\ensuremath{#1^{\circ}}}
\newcommand{\kms}{km\,s$^{-1}$}

\begin{document}

\title[Irradiation-driven warped discs]
  {SPH simulations of irradiation-driven warped accretion discs and the long periods in X-ray binaries}
\author[S. B. Foulkes et al.]
  {Stephen B. Foulkes$^{1}$, Carole A. Haswell$^{1}$, James R. Murray$^{2}$ \\
$^1$Department of Physics \& Astronomy, The Open University, Walton Hall, Milton Keynes, MK7 6AA, UK. \\
$^2$Department of Astrophysics \& Supercomputing, Swinburne University of Technology, Hawthorn, VIC 3122, Australia.\\
Email SBF: sbfoulkes@qinetiq.com, CAH: C.A.Haswell@open.ac.uk, JRM: jmurray@astro.swin.edu.au }

\date{Accepted. Received}
\pagerange{\pageref{firstpage}--\pageref{lastpage}}
\pubyear{2009}

\maketitle \label{firstpage}
\begin{abstract}
We present three dimensional smoothed particle hydrodynamics (SPH) calculations of
irradiation-driven warping of accretion discs.
Initially unwarped planar discs are unstable to
the radiation reaction when the disc is illuminated by a central radiation
source. The disc warps and tilts and
precesses slowly in a retrograde direction; its shape continuously
flexes in response to the changing orientation  of the Roche potential.
We simulate ten systems: eight X-ray
binaries, one cataclysmic variable (CV), and a `generic' low mass X-ray binary (LMXB).
We adopt system parameters from observations and tune a single parameter: our model X-ray luminosity
($L_{*}$) to reproduce the observed or inferred super-orbital periods.
Without exception, across a wide range of parameter space,
we find an astonishingly good match between
the observed $L_{X}$ and the model
$L_{*}$.
We conclude irradiation-driven warping is the mechanism
underlying the long periods in X-ray binaries.
Our Her\,X-1 simulation simultaneously reproduces
the observed $L_{X}$ , the ``main-" and ``short-high"
X-ray states and the orbital inclination.
Our simulations of SS\,433 give a maximum warp angle of
$18.6^{\circ}$, a good match to the cone traced
by the jets, but this angle is reached only in the outer disc.
In all cases, the overall disc tilt is less than $\degrees{13}$ and the maximum
disc warp is less than and or equal to $\degrees{21}$.
In particular, the disc warp in 4U\,1626-67 cannot
explain the observed torque reversals.
Taking our results at face value, ignoring the finite opening angle of the disc, we deduce orbital inclinations of approximately $\degrees{77}$ for 4U\,1916-053 and approximately $\degrees{69}$ for 4U\,1626-67.
We also
simulate Cyg\,X-2, SMC\,X-1, Cyg\,X-1,
and LMC\,X-3.
For high mass X-ray binary (HMXB) parameters, the discs' maximum angular elevation
is invariably at the outer edge.
For LMXBs with extreme mass ratios
a strong inner disc warp develops,
completely shadowing parts of the outer disc. This inner
warped disc executes retrograde precession while the outer disc
executes prograde apsidal precession.
The remaining LMXBs develop a less extreme warp in the inner disc,
with the entire disc tilting  and
precessing in a retrograde direction.
For our CV, KR\,Aur,
we matched the inferred disc precession period by adopting
$L_{X} = {\rm 10^{37}\,erg\,sec^{-1}}$,
which would require steady nuclear burning on the white dwarf surface.
\end{abstract}


\section{Introduction}
In \citet[][hereafter Paper I]{FoulkesEt:2006}, we presented three dimensional smoothed particle hydrodynamics (SPH) simulations of warped accretion discs in X-ray binary systems. We showed that geometrically thin, optically thick, initially
planar, accretion discs when irradiated by a central radiation source are unstable to warping. The central illumination of the non-axisymmetric disc surface causes a radiation pressure torque that acts on the disc to induce a twist or warp.
In general the warps precess in a retrograde direction.
Paper I discussed
X-ray binary systems with a range of mass ratios and luminosities.
Here we extend the work of Paper I, and  simulate the well-known
binary systems Hercules X-1 (Her\,X-1), SS\,433, 
several other X-ray binary systems, and one cataclysmic variable (CV). The X-ray binaries  modelled all have a super-orbital period, i.e. a period in the light curve that is much longer than the orbital period. 
We found for disc warping to occur the radiation intensity has to be greater than a critical value
\citep[c.f.][]{WijersPringle:1999}; for low radiation levels warping does not occur.

We use a non-linear SPH code based on that of
\cite{Murray:1996,Murray:1998} but completely rewritten and extended as described in Paper I.
Section \ref{HerX1:sec:Systems modelled} describes
the two best-known systems;
in sections \ref{sec:Basic Equations}, \ref{sec:Numerical method} and \ref{HerX1:sec:Simulations} we present our formulation of the problem using SPH; section \ref{HerX1:sec:Results} presents
and discusses our results; and section \ref{HerX1:sec:Conclusions} is devoted to conclusions.

\section{Systems modelled}
\label{HerX1:sec:Systems modelled}
In total we present models of ten systems
in this paper. Here in section~\ref{HerX1:sec:Systems modelled} we discuss in detail
Her\,X-1
and SS\,433, two archetypal warped disc X-ray binaries.
Her\,X-1
is bright and well-constrained observationally, so it is important for
assessment of our simulations.
SS\,433 has been rather less definitively characterised, and has kinematic
elements beyond the scope of our model: namely the relativistic jets and the accretion disc wind. Nontheless the model is an interesting counterpoint to that of Her\,X-1.
\subsection{Her\,X-1}
Her\,X-1 was discovered by \cite{TananbaumEt:1972}. It is an intermediate mass X-ray binary
consisting of an X-ray pulsar and
a 13th magnitude blue variable star HZ\,Her \citep{DavidsenEt:1972}. Sinusoidal variations and regular eclipses of the X-ray source indicates the system is viewed at a high inclination with an orbital period of $1.70$ days. Assuming HZ\,Her fills its Roche lobe and using the eclipse duration \cite{TananbaumEt:1972} estimated the mass ratio of the system to be $M_{\rm donor} / M_{\rm compact} \approx 1.9$.
In Table \ref{HerX1:table:HerX1Parameters} we summarise  the parameters we have adopted  for Her\,X-1.
\begin{table}
\caption
  {
    Major parameters for binary system HZ\,HER/Her\,X-1
  }
\label{HerX1:table:HerX1Parameters}
\begin{center}
\begin{tabular}{l l l}
\hline
Parameter                      & value                                          & reference    \\
\hline
\hline
orbital period                 & $P_{\rm orb}$  =  1.7 days                        & 1     \\
binary separation              & $a$        = ${\rm 6\cdot10^{11}}${\rm cm}                 & 2     \\
orbit eccentricity             & $e$        = ${\rm 3\cdot10^{-4}}$                   & 2     \\
inclination of system          & $i$        = \degrees{80} - \degrees{90}       & 2,3,4 \\
mass of neutron star           & $M_{1}$   = ${\rm 1.3 M_{\odot}}$                   & 2,3   \\
mass of donor                  & $M_{2}$   = ${\rm 2.2 M_{\odot}}$                   & 2,3   \\
mass transfer rate    & $-\dot M_2$ = ${\rm 10^{17.9}}$ g sec$^{-1}$             & 5     \\
X-ray luminosity               & $L$        = ${\rm 10^{37.3}\,erg\,sec^{-1}}$          & 6     \\
super-orbital period           & $P_{\rm prec}$ = ${\rm 34.88\pm0.12}$     {\rm days}               & 9     \\
disc radius                    & $r_{\rm disc}$      = ${\rm 1-2\cdot 10^{11}\,{\rm cm}}$              & 10,11,12 \\
\hline
\end{tabular}
\end{center}
References: $^1$ \cite{TananbaumEt:1972}, $^2$ \cite{DeeterEt:1981}, $^3$ \cite{MiddleditchNelson:1976}, $^4$ \cite{GerendBoynton:1976}, $^5$ \cite{Schandl:1996}, $^6$ \cite{McCrayEt:1982}, $^7$ \cite{TruemperEt:1978}, $^8$ \cite{PravdoEt:1977}, $^9$ \cite{GiacconiEt:1973}, $^{10}$ \cite{Oke:1976},  $^{11}$ \cite{CrosaBoynton:1980}, $^{12}$ \cite{Middleditch:1983}
\end{table}

\cite{TananbaumEt:1972} and  \cite{GiacconiEt:1973}
reported a super-orbital X-ray cycle
due to periodic obscurations of the X-ray source by the disc material.
The $34.88\pm0.12$ day cycle is comprised of
on and off states lasting for 11-12 and 24 days respectively.
\cite{JonesForman:1976} observed during the $24$ day off state and discovered
the ``short-high" emission which is approximately $30\%$ of the peak intensity of the ``main-high".

\cite{GiacconiEt:1973} also reported regular dips in the X-ray light curve, the duration of which was about $1-3$ hours. The frequency of the dips was related to the sum of frequencies of the system orbital and the super-orbital frequency. They reported two types of dips. ``Preeclipse" dips which appear just before an eclipse, at orbital phase $(\phi_{orb})$ approximately $0.8$, which propagate back in super-orbital phase as the $35$ day phase increases. These dips recur on a period of approximately $1.65$ days. The second set of dips are known as the ``anomalous dips". These dips occur at $\phi_{orb} \approx 0.5$ and are observed during the ``short-high". The X-ray dips can be explained by a number of different models, but they are certainly produced by occultation of the central X-ray source by the accreting disc material.

It is generally accepted that the 35 day super-orbital
period is a consequence of a twisted/warped \citep {Kumar:1986,Kumar:1990} and tilted \citep {Roberts:1974} accretion disc that precesses in a retrograde direction about the central neutron star \citep {Petterson:1975,Petterson:1977,Petterson:1978}. An observer sees the disc nearly edge on and as regions of the disc move above the orbital plane by more than the disc inclination the central X-ray source will be occulted.  \cite{Katz:1973} proposed that the tidal force of HZ Her acts on the tilted accretion disc and causes it to precess, and \cite{LarwoodEt:1996} find this mechanism produces the correct precession rate.
Neither of these studies, however, explain the origin of the tilt and twist of the disc.

\cite{SchandlMeyer:1994} and \cite{Schandl:1996} used a coronal disc wind model to generate a warped disc through asymmetric wind induced force on the disc surface. They modelled an optically thick, geometrically thin accretion disc that was heated by X-rays originating from the neutron star, which resulted in a hot corona above the disc. Near the disc the corona was hydrostatically layered and optically thick. However, in the outer regions the gas exceeds the escape velocity and leaves the gravitation potential of the binary system. This coronal wind
generates a reaction force on the disc surface, which results in the outer disc regions warping and the inner parts being tilted but unwarped.
For an orbital  inclination of \degrees{80} they were able to reproduce the ``main-high" and ``short-high". They also used a simple model to generate the observed X-ray dips including the back propagation of the dips as the super-orbital precession proceeds.

\subsection{SS\,433}
SS\,433 is an unusual, well-studied binary system
associated with the supernova remnant W50 \cite{Clark:1984};
see \cite{Margon:1984} and \cite{Fabrika:2006} for reviews.
Intensive observations led to a model that includes an evolved binary
with a large mass transfer rate and $P_{\rm orb}$ approximately $13$ days.
A Roche lobe-filling donor star loses
mass to a black hole \citep{Blundell:2008} via an accretion disc.
Some of the inflow is directed through the interaction of the disc and the compact object into oppositely-directed relativistic precessing jets.

``Moving"
emission lines have
extreme radial velocities with
a period of approximately $162$ days.
They
are described by the ``kinematical model" of the
jets driven
by a precessing  accretion disc \citep{FabianRees:1979,
Milgrom:1979,AbellMargon:1979}.
The ``moving'' lines are also seen in the Chandra X-ray spectra \citep{MarshallEt:2002}.
Another major periodicity found is the so-called ``nodding" of the disc \citep{KatzEt:1982}. This is a $6.28$ day period and is attributed to the donor star's
passage through the nodal line of the titled disc
which causes
tidal deformation of the disc. In addition to the jets there is evidence of a very powerful outflow from the disc with a velocity of approximately $1500$ \kms \citep[see][and references therein]{BegelmanEt:2006}. Figure 1a of \cite{BarnesEt:2006} shows the major elements of SS\,433, which include the compact object, the Roche filling donor star, the accretion disc, the precessing relativistic jets and the outflow in a form of extend disc or disc wind.

The precessional and orbital periods have provided a very satisfactory model of the motion of the jets \citep{KatzEt:1982,CollinsGarasi:1994} and the photometric variations \citep{GoranskiiEt:1998}. However, there are still many unanswered questions about the system. These include, the
formation process for the jets and the disc precession
mechanism \citep{WijersPringle:1999}.
In Table \ref{HerX1:table:SS433Parameters} we give a summary of the main parameters we adopted for SS\,433.
\begin{table}
\caption
  {
    Major parameters adopted for binary system SS\,433
  }
\label{HerX1:table:SS433Parameters}
\begin{center}
\begin{tabular}{ l l l }
\hline
Parameter                      & value                                & reference    \\
\hline
\hline
orbital period                 & $P_{\rm orb}$  = $13.1$ days             & 1 \\
inclination of system          & $i$        = $\degrees{78.83}$       & 1 \\
system mass ratio              & $q$        = $2$                     & 3$^\dagger$ \\
mass of compact star           & $M_{1}$    = $10\,M_{\odot}$          & 3 \\
mass of donor                  & $M_{2}$    = $20\,M_{\odot}$          & 3 \\
mass transfer rate & $-\dot M_2$ = $10^{-3}M_{\odot}$ yr$^{-1}$   &  6    \\
X-ray luminosity               & $L$        = $10^{36}{\rm \, erg\, sec^{-1}}$  & 2, 4, 5 \\
super-orbital period           & $P_{prec}$ = $162.5$ days            & 1 \\
disc nodding                   & $P_{nod}$  = $6.2877$ days           & 1 \\
disc tilt  & {\rm disc tilt}   $\approx\degrees{20}$         & 1 \\
\hline
\end{tabular}
\end{center}
$^{\dagger}$ We note the recent paper \cite{Blundell:2008} favours $M_{1} = 16 M_{\odot}$, $M_{2} = 22 M_{\odot}$, and consequently $q=1.38$. The simulation reported here predated this finding.

References: $^1$\cite{MargonAnderson:1989}, $^2$\cite{BrinkmannEt:1991}, $^3$\cite{WijersPringle:1999}, $^4$\cite{KotaniEt:1996}, $^5$\cite{MarshallEt:2002}, $^6$\cite{Shklovskii:1981}.
\end{table}

\section{Basic Equations}
\label{sec:Basic Equations}
We take the basic hydrodynamics equations governing the viscous gaseous material of the accretion disc. It is assumed that the disc self-gravity is negligible in comparison to the binary component's tidal field. The equations of continuity and motion may be written as

\begin{equation}\label{eq:1}
\frac {D \rho}{D t} + \rho \mathbf{\nabla} \cdot \mathbf{v} = 0
\end{equation}

\begin{equation}\label{eq:2}
\frac {D \mathbf{v}}{D t} =  -\frac {1} {\rho} \mathbf{\nabla} P - \mathbf{\nabla} \Phi + \mathbf{f_{v}} + \mathbf{f_{irr}}
\end{equation}

\noindent where in the above

\begin{equation}\label{eq:2a}
\frac {D }{D t} \equiv  \frac {\partial}{\partial t} + \mathbf{v} \cdot \mathbf{\nabla}
\end{equation}  

\noindent is the material derivative operator \citep{Batchelor:1970}, $\rho$, $\mathbf{v}$ and $P$ are the fluid density, velocity and pressure values respectively. The gravitational potential is $\Phi$, the viscous force is $\mathbf{f_{v}}$ and the force due to the radiation on the fluid is $\mathbf{f_{irr}}$. 
The equation for the rate of change of thermal energy per unit mass can be expressed as
\begin{equation}\label{eq:2c}
\frac {d u}{d t} =  - \left ( \frac {P}{\rho} \right ) \mathbf{\nabla} \cdot \mathbf{v}.
\end{equation}  

\section{Numerical method}
\label{sec:Numerical method}
We solve the basic equations in section \ref{sec:Basic Equations} using a SPH \citep{Lucy:1977, GingoldMonaghan:1977} computer code first developed by Murray (1996, 1998).

\subsection{SPH formulation}
SPH uses smoothed quantities defined by the equation
\begin{equation}\label{eq:3}
\langle f\left (\mathbf{r}\right) \rangle = \int f(\mathbf{r'})W\left( \vert \mathbf{r-r'}\vert, h\right )d\mathbf{r'},
\end{equation}  

\noindent here $f(\mathbf{r})$ is an arbitrary function, $W$ is a smoothing kernel, $h$ is the smoothing length and the integration is over the entire space.  A Monte-Carlo representation of $(\ref{eq:3})$ is used for each particle position

\begin{equation}\label{eq:4}
\langle f\left (\mathbf{r_{i}}\right) \rangle = \sum_{j=1}^N \frac{m_{j}}{\rho_{j}}f(\mathbf{r_{j}})W\left( \vert \mathbf{r_i-r_j}\vert, h\right ),
\end{equation} 

\noindent where $m_j$ is the mass of particle $j$, $\rho_j$ is the density of the particle $j$ and the sum is over all particles N. We use a variable smoothing length such that each particle, $i$, has an associated smoothing length $h_i$. The kernel is symmetrized  with respect to particle pairs in order to conserve momentum.

\subsubsection{Equations of motion}
The equation of motion for each particle $i$ is given by
\textsl{}
\begin{equation}\label{eq:5}
m_i\frac{d\mathbf{v}_i}{dt}=\mathbf{F}_{P,i}+\mathbf{F}_{G,i}+\mathbf{F}_{visc,i}+\mathbf{F}_{irr,i},
\end{equation} 

\noindent which gives the force on particle $i$ as a summation of the pressure, gravity, viscous and irradiation forces, $\mathbf{F}_{P,i}$, $\mathbf{F}_{G,i}$, $\mathbf{F}_{visc,i}$ and $\mathbf{F}_{irr,i}$ respectively.

\subsubsection{The pressure term}
For the pressure gradient we make use of the fact that

\begin{equation}\label{eq:6}
\frac{1}{\rho}\mathbf{\nabla} P = \mathbf{\nabla} \left ( \frac{P}{\rho} \right ) + \frac{P}{\rho^2} \mathbf{\nabla} \rho,
\end{equation} 

\noindent and the pressure is given by

\begin{equation}\label{eq:7}
\left ( \frac{1}{\rho}\mathbf{\nabla}P \right )_i = \sum_{j} m_j \left (\frac{P_i}{\rho_i^2} + \frac{P_j}{\rho_j^2} \right) \mathbf{\nabla}W \left(\vert \mathbf{r_i} - \mathbf{r_j} \vert, h \right ) .
\end{equation} 

\noindent Similarly the SPH energy equation is 
\begin{equation}\label{eq:7a}
\frac{du_i}{dt} = \frac{1}{2} \sum_{j} m_j \left (\frac{P_i}{\rho_i^2} + \frac{P_j}{\rho_j^2} \right)\mathbf{v_{ij}} \cdot \mathbf{\nabla} W \left(\vert \mathbf{r_i} - \mathbf{r_j} \vert, h \right ) .
\end{equation} 

\subsubsection{The artificial viscosity term}
To stabilise the shock flow regions we used an artificial viscosity term as detailed by \cite{Murray:1996} and \cite{TrussEt:2000}. An artificial viscosity term was added to the pressure term such that

\begin{equation}\label{eq:8}
\frac{P_i}{\rho_i^2} + \frac{P_j}{\rho_j^2} \rightarrow \frac{P_i}{\rho_i^2} + \frac{P_j}{\rho_j^2} + \Pi_{ij}
\end{equation} 

\noindent where the artificial viscous pressure is given by the \cite {GingoldMonaghan:1983} equation

\begin{equation}\label{eq:9}
\Pi_{ij} = \frac{1}{\bar\rho_{ij}}\left(-\alpha\mu_{ij}\bar c_{ij}  + \beta\mu_{ij}^2 \right).
\end{equation} 

\noindent Here we are using the notation $\bar A_{ij} = \frac{1}{2}\left(A_i+A_j\right)$ and $\bar c_{ij}$ is the average sound speed at particle positions $i$, $j$ and


\begin{equation}\label{eq:10}
\mu_{ij} = \bar h_{ij}\frac{\mathbf{v_{ij}}\cdot \mathbf{r}_{ij}}{\mathbf{r}_{ij}^2+\eta^2}. 
\end{equation} 

\noindent In this equation $\mathbf{v}_{ij} = \mathbf{v}_i-\mathbf{v}_j$ and $\eta^2=0.01\bar h_{ij}^2$ which prevents the denominator going to zero. We used $\alpha = 0.5$ and $\beta=0$ for the simulations reported here.  

\subsubsection{The radiation force}
 We follow the geometrical definitions of \cite{Pringle:1996} and define a unit tilt vector (see Figure 1 of \cite{Ogilvie:1999}) $\mathbf{l}(R,t)$ such that

\begin{equation}\label{eq:11}
\mathbf{l}=(\cos\gamma \sin\beta, \sin\gamma \sin \beta, \cos \beta)
\end{equation} 

\noindent where $\beta(R,t)$ and $\gamma(R,t)$ are the Euler angles of the tilt vector with respect to a fixed Cartesian coordinate system $(OXYZ)$ centered on the compact object.
The position vector of a point on the disc with distance $R$ from the origin and azimuth $\phi$ is given by $R\mathbf{e_{R}}$ where $\mathbf{e_{R}}$ is the radial unit vector and is given by

\begin{equation}\label{eq:12}
\mathbf{e_{R}}=\left[ \begin{array}{c}
\cos\phi \sin\gamma + \sin\phi \cos\gamma \cos \beta \\
-\cos\phi \cos\gamma + \sin\phi \sin\gamma \cos\beta \\
-\sin\phi \cos\beta 
\end{array} \right] .
\end{equation} 

\noindent The radiation source is assumed to be centered on the compact object position and radiates isotropically. The radiation flux at a distance $R$ from the compact object is given by

\begin{equation}\label{eq:13}
\mathbf{f}=\frac{L_{*}}{4\pi R^2} \mathbf{e_{R}},
\end{equation} 
and the power absorbed at a surface element $d\mathbf{S}$ is

\begin{equation}\label{eq:14}
dP= \left (1-\eta \right ) \big | \mathbf{f} \cdot d\mathbf{S} \big | ,
\end{equation} 

\noindent where $L_{*}$ is the total luminosity of the radiation source, $\eta$ is the disc albedo, that is the amount of radiation scattered by the surface of the disc, in all the simulations reported in this paper $\eta$ was assumed to be zero, and $d\mathbf{S}$ is an element of the disc surface area given by

\begin{equation}\label{eq:15}
d\mathbf{S}=\left(\frac{\partial \mathbf{R}}{\partial R}dR \right) \times \left(\frac{\partial \mathbf{R}}{\partial \phi}d\phi \right).
\end{equation} 

\noindent We assume that the radiation is absorbed by this element and is uniformly re-radiated immediately and on the same disc side. The force due to absorption of the radiation is

\begin{equation}\label{eq:16}
d\mathbf{A}= \frac{dP}{c} \mathbf{e_R},
\end{equation} 

\noindent where $c$ is the speed of light. 

Since the incident radiation is parallel to the radius vector a circular disc annulus receives no torque from the absorbed radiation. There is, however,  a net torque on a circular annulus due to the summed radiation pressure reaction, as calculated by \cite{Pringle:1996}.  Since the radiation is re-radiated uniformly from the disc surface, the element will receive a radiation pressure reaction from this radiation of the form

\begin{equation}\label{eq:17}
d\mathbf{F}= -\frac{2}{3}\frac{dP}{c}\mathbf{n},
\end{equation} 

\noindent where $\mathbf{n}$ is the unit normal pointing away from the disc surface that received the initial radiation.

\section{Simulations}
\label{HerX1:sec:Simulations}
The results presented here were generated using a local private Linux system and The Swinburne Centre for Astrophysics and Supercomputing facility.
%

In the simulations the total system mass, $M_{t}$, and the binary separation, $a$, were both scaled to unity. The binary orbital period, $P_{orb}$, was scaled to $2\pi$. The radiation source was also scaled such that

\begin{equation}\label{eq:18}
\frac{L_{*Physical}}{L_{*Code}}=G^{2/3} \left( \frac{2 \pi M_{t}}{P_{orb}}\right)^{5/3}.
\end{equation} 

In the simulations the accretion disc had an open inner boundary condition in the form of a hole of radius $r_{1}=0.0125a$ centred on the position of the primary object. Particles entering the hole were removed from the simulation. Particles that re-entered the secondary Roche lobe were also removed from the simulation as were particles that were ejected from the disc and had a distance $> 0.9a$ from the centre of mass of the system.

We assumed a local isothermal equation of state and that the dissipation was radiated promptly from the point at which it was generated.  The \cite{ShakuraSunyaev:1973} viscosity parameters were set to $\alpha_{low}$ of 0.01 and $\alpha_{high}$ of 0.10, and the viscosity state changed smoothly as described in \cite{TrussEt:2000}. The SPH smoothing length, $h$, was allowed to vary in both space and time and had a maximum value of $0.01a$.

\subsubsection{The gas stream}
We simulated the mass loss from the secondary by introducing particles at the inner Lagrangian point ($L_1$). See Paper I for a description of the particle initial velocities.

\subsubsection{The initial non-warped accretion disc}

The simulations were started with zero mass in the accretion disc and with the radiation source switched off. A single particle was injected into the simulation every $0.01\Omega_{orb}^{-1}$ at the $L_1$ point until a quasi-steady mass equilibrium was reached within the disc. The simulations were then continued for another $3$ orbital periods to ensure mass equilibrium. The number of particles in each accretion disc was approximately $30,000$ giving a good spatial resolution. All the systems modelled had moderate vertical resolution of the order 3-5 smoothing lengths. The radiation source intensity was turned on gradually using a time exponential that increased the power of the radiation over a short period of time.

\subsection{Surface finding algorithm \& self-shadowing}

The accretion disc surface finding algorithm has been described in Paper I. Basically, the method uses a convex hull algorithm to find the surface particles. The accretion disc surface was then constructed from the set of surface points that form a convex hull of the disc particles. We used a ray-tracing algorithm  to determine regions of self-shadowing. For each particle found on the disc surface a light-ray was projected from the particle to the position of the radiation source at the centre of the disc. The particle was deemed to be illuminated by the radiation source if this light-ray did not intersect any disc material between the particle surface position and the radiation source. The radiation force was only applied to particles that were considered to form part of the disc surface and were illuminated by the central radiation source.

\subsection{Disc warping and precession measure}

The two measures defined by \cite{LarwoodPapaloizou:1997} were used to measure the disc warping and the amount of warp precession. They defined an angle $j$ as the angle between the total disc angular momentum vector and the angular momentum vector for a specific disc annulus, i.e.

\begin{equation}\label{WapredDiscs:eq:22}
\cos\ j = \frac{\mathbf{J}_A \cdot \mathbf{J}_D}{\big | \mathbf{J}_A \big | \big | \mathbf{J}_D \big |}.
\end{equation}

\noindent The term $\mathbf{J}_A$ is the total angular momentum within the specific annulus and was calculated by summing the angular momentum for each particle within the annulus. The term $\mathbf{J}_D$ is the total disc angular momentum and was calculated by summing all the angular momenta for all particles within the disc. An angle $\Pi$ was also defined which measures the amount of precession of the disc angular momentum relative to the initial binary orbital angular momentum, $\mathbf{J}_O$:

\begin{equation}\label{WapredDiscs:eq:23}
\cos\ \Pi = \frac{\left(\mathbf{J}_O \times \mathbf{J}_D\right) \cdot \mathbf{u}}
{\big | \mathbf{J}_O \times \mathbf{J}_D \big | \big | \mathbf{u} \big |}
\end{equation}

\noindent where $\mathbf{u}$ is any arbitrary vector in the binary orbital plane. We also measured the tilt of the entire warped accretion disc relative to the orbital plane using
\begin{equation}\label{WapredDiscs:eq:24}
cos\ \beta = \frac{\mathbf{J}_O \cdot \mathbf{J}_D } {\big | \mathbf{J}_O \big | \big | \mathbf{J}_D \big | }.
\end{equation}

\subsection{Parameters Adopted and Adjusted}
\label{HerX1:sec:Methods}
For each simulation we adopted system parameters from the literature, as summarised in Tables 1, 2, 3 and 4.
We built up the disc to mass equilibrium before turning on the radiation source.
The precession rate in the simulations was dependent on the
radiation intensity, as
described in Paper I; by adjusting the value of $L_{*}$ we were able to reproduce the observed disc precession rates. {\bf We emphasize that the ONLY parameter we adjusted to match the published precession rate was the luminosity of the central source}, all other input parameters were adopted from the literature before the simulations were run.

\section{Simulation Results}
\label{HerX1:sec:Results}

\subsection{Her\,X-1}
The SPH accretion disc model for the Her\,X-1 simulation without the radiation source was roughly point-symmetric about the compact object. The disc did not precess and remained approximately constant in shape and size once mass equilibrium had been reached.

\begin{figure*}
  \centerline{\epsfig{file=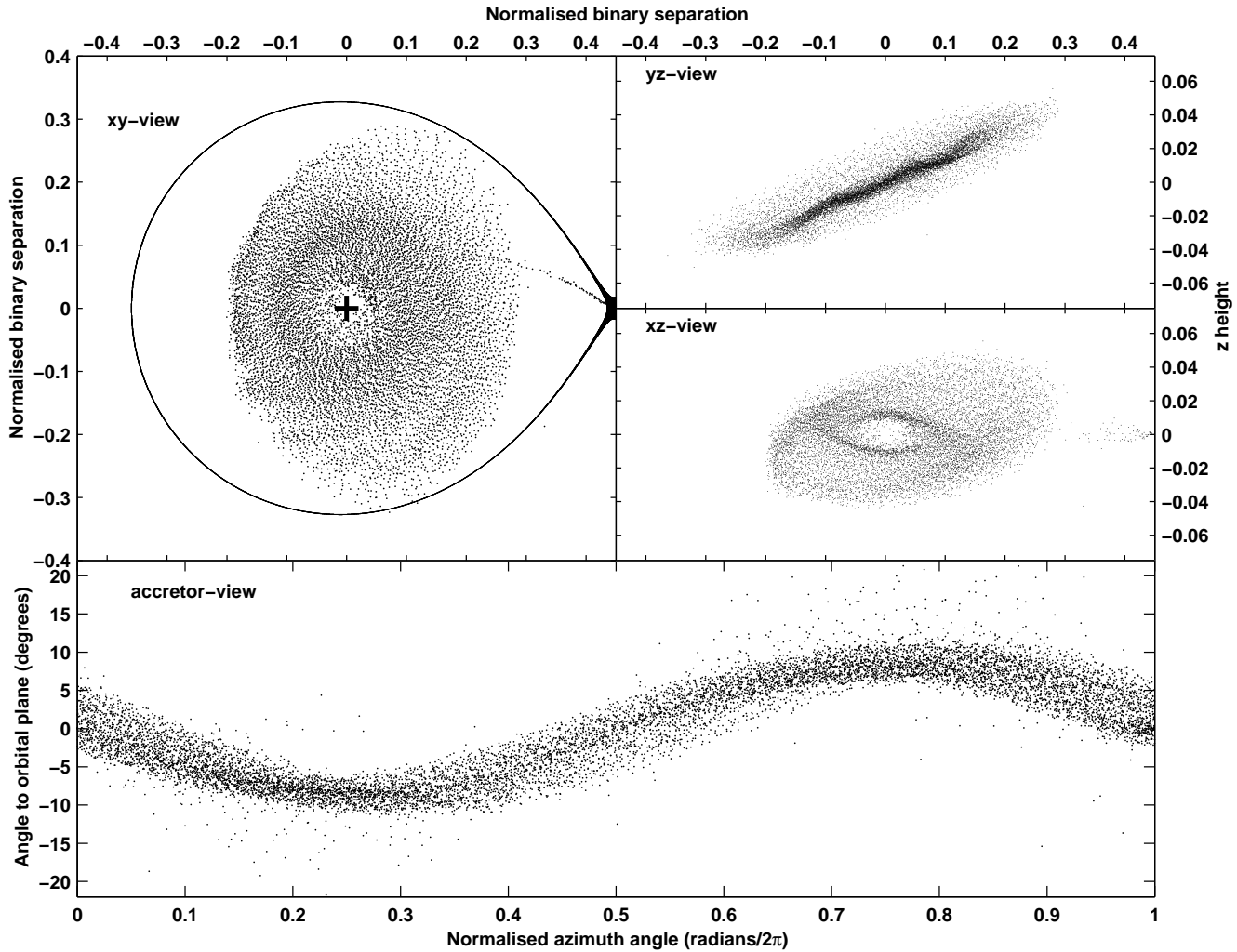,width=1.0\textwidth,angle=0}}
  \caption{
	         Particle projection plots for Her\,X-1 simulation. The position of each particle is indicated by a small black dot. The plot labelled  xy-view is a plan view of the accretion disc as seen from above the disc. The cross at the centre of the plot shows the position of the primary object. The solid dark line is the Roche lobe of the primary and the $L_1$ point is to the right of the plot. The two plots xz-view and yz-view are particle projection plots on a plane perpendicular to the orbital plane and through the system axis. The bottom plot, accretor-view, shows the particle distribution as seen from Her\,X-1. The horizontal axis is the normalised azimuth angle, the $L_{1}$ point is at azimuth angle 0 and the stream/disc impact region is at an angle of approximately $0.9$. The vertical axis is the angle, in degrees, between a particle and the orbital plane when viewed from the compact object. The disc material flows from right to left.
         }
  \label{HerX1:Figure:herx1_part_pos}
\end{figure*}

When the radiation source was switched on a strong twist developed in the disc and the entire disc tilted out of the orbital plane. Figure \ref{HerX1:Figure:herx1_part_pos} contains projection plots for the Her\,X-1 model; the position of each SPH particle is indicated by a small black dot.  The upper left hand plot, labelled xy-view, is a plan view of the accretion disc as seen from above the disc. The solid dark line is the Roche lobe of the primary and the $L_1$ point is to the right of the plot. Material from the secondary enters the primary Roche potential from the $L_1$ point. The cross at the centre of the map is the position of the primary object. The disc remained approximately constant in shape and size throughout the simulation.

The two upper right-hand plots of Figure \ref{HerX1:Figure:herx1_part_pos}, labelled yz-view and xz-view, are side views of the disc in the y-z and x-z directions respectively. The yz-view plot is a projection view of the disc as seen from the secondary, similarly the xz-view is a projection plot with the secondary located to the right of the plot.
The lower plot of Figure \ref{HerX1:Figure:herx1_part_pos}, labelled accretor-view, shows the distribution of the particles as seen from the compact object.
The disc tilt out of the orbital plane is clearly seen in the yz-view plot,
where
the disc is viewed almost side on. The inner disc warp can be seen on
either side of the position of the accretor. This warp is odd symmetrical about the centre of the disc. The maximum value of the warp is located at a distance of approximately $0.1a$ either side of the primary position.

Figure~\ref{HerX1:Figure:herx1_part_pos}
shows
that the radiation force has pushed the entire disc out of the orbital plane. The inclination of the disc remained constant for many orbital periods and the disc precessed in a retrograde direction relative to the orbital flow.

\begin{figure*}
  \centerline{\epsfig{file=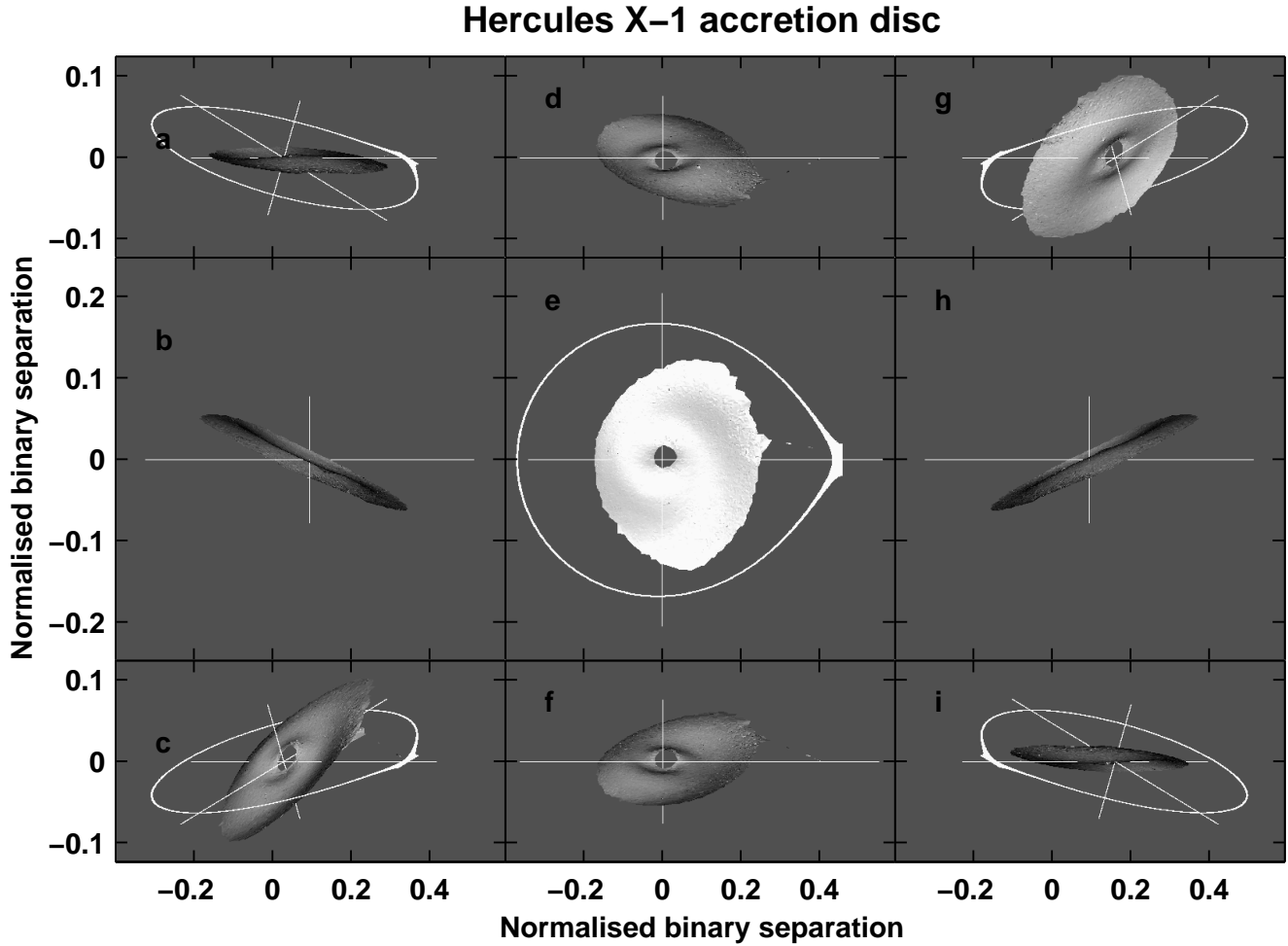,width=1.0\textwidth,angle=0}}
  \caption{
	         Her\,X-1 SPH accretion disc viewed at different angles.
Note the direction perpendicular to the orbital plane has been artificially
magnified by a factor of two.
This renders the vertical structure more visible, but causes the tilt angles
to appear much larger than they really are.
The curved white line is the Roche lobe of the compact object and the straight white lines form axes centred on the compact object.
The disc is illuminated using an external white light source place along the z-axis.
         }
  \label{HerX1:Figure:herx1_geomview}
\end{figure*}

In Figure \ref{HerX1:Figure:herx1_geomview} we show the shape of the disc viewed at different orientations. Note the z-scale, i.e. the direction
perpendicular to the orbital plane, is magnified by a factor of two,
making the vertical structure more visible.
The tilt angle visually inferred from Figure~\ref{HerX1:Figure:herx1_geomview} is
therefore misleading.
The disc is illuminated by an external light source that lies in the z-direction. The curved white line in each plot is the Roche lobe of the primary object and the straight white lines indicate the xyz axes defined above.
The central image (e) shows the accretion disc as viewed from above, the disc shape and warp is clearly seen in this image. The images (b), (d), (f) and (h) are views of the disc from the sides, back and front of the disc. The remaining images (a), (c), (g) and (i) are views of the disc from slightly above or below the orbital plane.

\begin{figure}
  \centerline{\epsfig{file=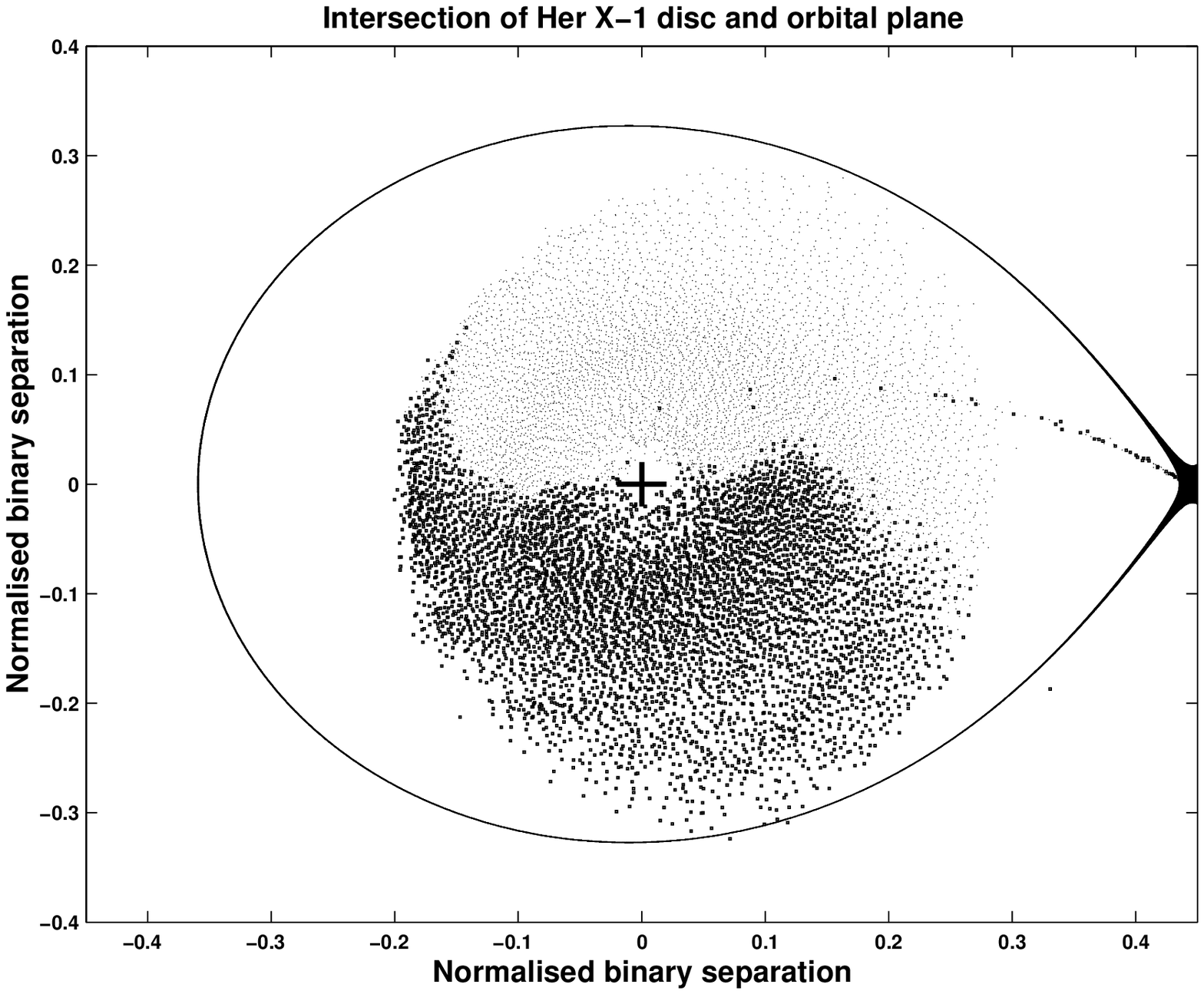,width=0.5\textwidth,angle=0}}
  \caption{
	         Her\,X-1 particle plot plan view. The particles above the orbital plane are indicated using a small square that appear dark in the plot. The remaining particles, that are below or on the orbital plane are indicated using a small dot. The boundary of the two sets of points shows the intersection of the accretion disc and the orbital plane.
         }
  \label{HerX1:Figure:herx1_disc_plane}
\end{figure}

Figure \ref{HerX1:Figure:herx1_disc_plane} is similar
to the xy-view in Figure~\ref{HerX1:Figure:herx1_part_pos}. Here
particles that lie above the orbital plane are indicated by a small square that appear dark on the plot. All other particles are indicated using a small dot. The boundary between the two sets of points shows the Eulerian angle of the warp precession, $\gamma$. For this snapshot the Eulerian angle, $\gamma$, is approximately odd symmetrical with respect to the y-axis. This boundary line describes the intersection between the warped accretion disc and the orbital plane. As the disc precesses in a retrograde direction the position of the Eulerian angle, $\gamma$, also rotates at the same rate.

\begin{figure}
  \centerline{\epsfig{file=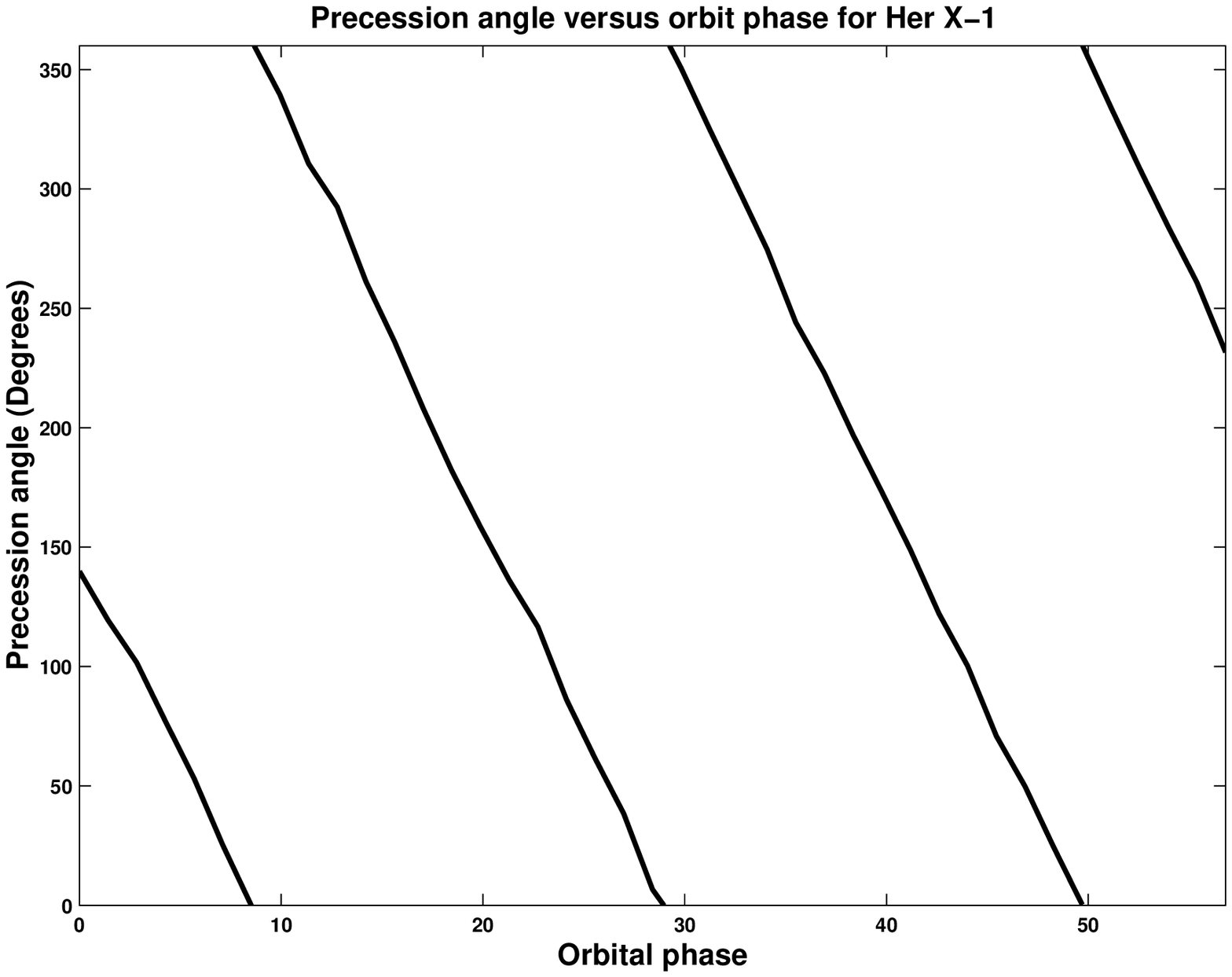,width=0.5\textwidth,angle=0}}
  \caption{
	         Disc precession angle versus orbital phase for the Her\,X-1 SPH simulation. The graph shows the direction of the angular momentum for the whole disc measured relative to some arbitrary starting position. The disc rotates in a retrograde direction with a precession period of approximately $34.9$ days.
         }
  \label{HerX1:Figure:herx1_prec_plot}
\end{figure}

The disc precessed in a retrograde direction relative to the
inertial frame, the entire disc precessing with a single period. This precession
is accompanied by continuous flexing of the disc structure as
its orientation with respect to the Roche potential changes due to the
orbital motion of the donor star.
We tuned the luminosity to a value of
$L_* = 2.15 \times 10^{37} {\rm erg\thinspace s^{-1}}$ to achieve
the observed precession
rate of approximately $34.9$ days.
This value of $L_*$ is approximately $10\%$ higher than the value quoted in
Table 2. This discrepancy is, of course, well within the uncertainty
arising from the distance adopted by
\citet{McCrayEt:1982}.
More importantly,
the factor implicitly introduced by our assumption of zero
albedo is probably approximately $2$, i.e. far larger than approximately $10\%$.
The warp shape was very stable over many orbital periods; we ran this simulation for more than 60 orbital periods and the warp shape remained constant throughout. The rate of precession of the disc was determined as described in paper I.
Figure \ref{HerX1:Figure:herx1_prec_plot} shows the precession angle versus orbital phase for the model.
The disc precession rate was determined by fitting a straight line to the data using a
Numerical Recipes least squares method \citep{PressEt:1986}.

The warped disc is inclined relative to the orbital plane as shown in Figure \ref{HerX1:Figure:herx1_part_pos} (yz-view) which changes the point at which the gas stream impacts the accretion disc. For a planar
accretion disc the material from the gas stream always impacts the outer edge of the disc at approximately the same point and the energy dissipated at this point is approximately constant with time. This is not true for a warped disc. As the secondary orbits around
the warped, tilted disc, the impact point of the gas stream moves from the
edge of the disc to nearer its centre.
Consequently the kinetic energy gained by the ballistic stream, and subsequently
dissipated at the impact, varies.

\begin{figure*}
  \centerline{\epsfig{file=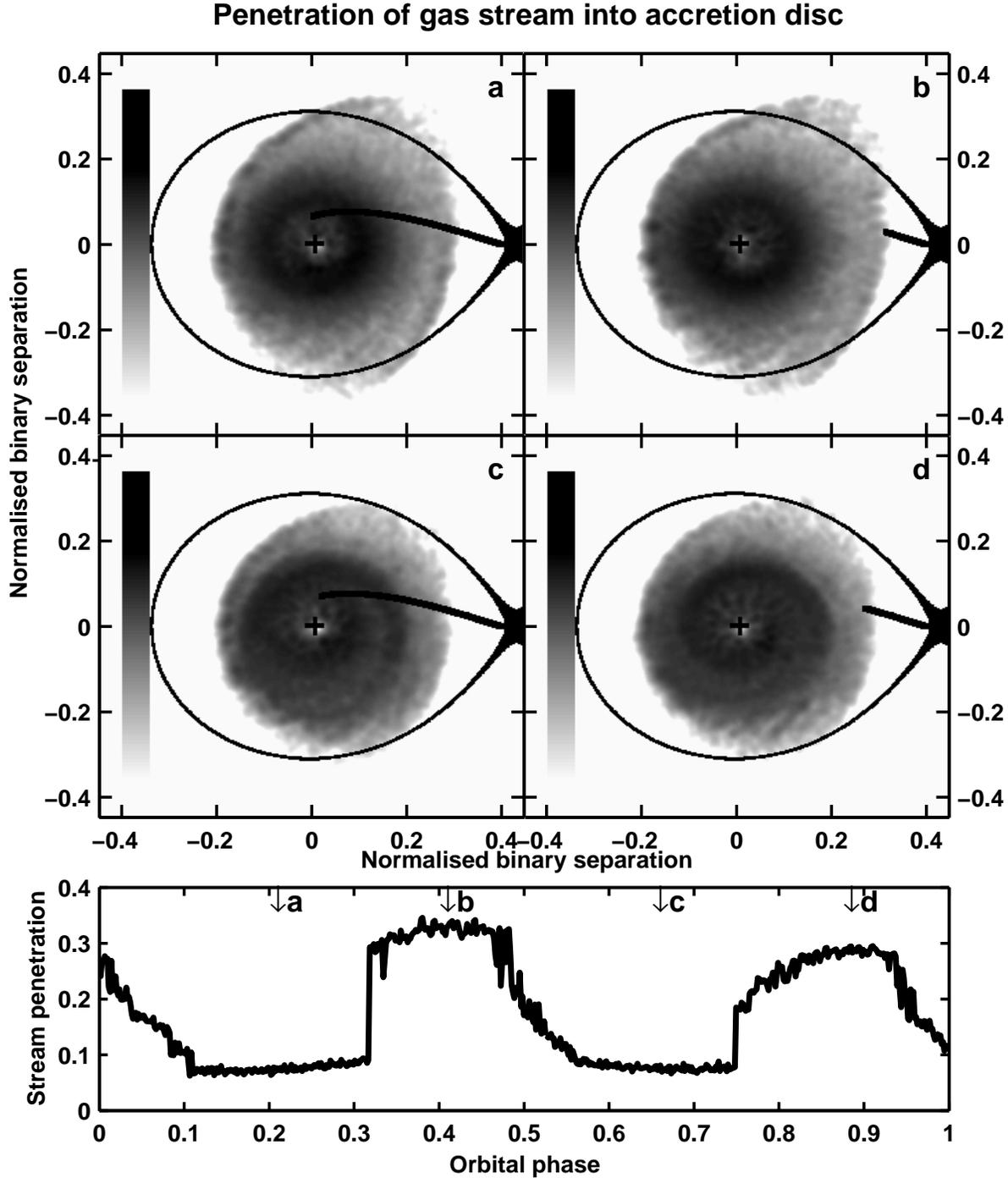,width=0.9\textwidth,angle=0}}
  \caption{
	         Penetration of the gas stream for the Her\,X-1 simulation. The lower plot shows the stream/disc impact distance from the neutron star as function of orbital phase. The four upper plots are density maps of the accretion disc as viewed from above of the disc. The Roche lobe of the neutron star is also plotted. The ballistic path of the gas stream is plotted, it starts at the $L_{1}$ point and ends when the gas stream intercepts the accretion disc.
         }
  \label{HerX1:Figure:herx1_stream_disc}
\end{figure*}

The lower plot of Figure \ref{HerX1:Figure:herx1_stream_disc} shows how
the distance from the compact object of the
stream/disc impact point varies with the orbital phase.
This plot does not repeat with the orbital phase, instead it repeats with a period of 1.65 days which is the synodic period between
the orbital period and the retrograde  precession period of
the accretion disc:
$P_{syn}^{-1} = P_{orb}^{-1} + P_{prec}^{-1}$.

The upper plots of Figure \ref{HerX1:Figure:herx1_stream_disc} (a, b, c and d) show disc density maps and the ballistic trajectory of the gas stream from the $L_{1}$ point to the impact point on the accretion disc. The orbital phases of plots (a, b, c and d) are indicated on the lower plot. Plots (a) and (c) show that the gas stream impacts the disc very close to the central object. The density maps also show that the disc is eccentric and changing shape on the orbital
period. The tidal action of the secondary is making the disc flex and distort.
Tidally-driven flexing will occur whenever the disc is not fixed within the
primary Roche lobe; see \citet{HaswellEt:2001} and \citet{Smithench:2007} for
a discussion of this for apsidally precessing discs.

In the interval 
between orbital phases approximately $0.11$ and approximately $0.31$ the gas stream flows over the disc surface
and impacts the disc very close to the central object (see Figure~\ref{HerX1:Figure:herx1_stream_disc}). At
orbital phase approximately $0.31$ the impact point very abruptly changes to
near the disc edge.
The stream continues to interact with the edge of the disc for approximately $0.18 P_{orb}$. The gas stream then interacts with the other surface
of the disc
and slowly
the warp and tilt opens allowing the gas stream to flow further towards the
central object. There is then a second abrupt change of the impact
distance at about orbital phase approximately $0.75$ as the gas stream again impacts close to the edge of the disc. This change is not as large as the first because the disc shape is asymmetric.
The gas stream interacts with the disc near the edge of the disc for approximately $0.16$ $P_{orb}$ and then slowly moves closer to the central object on the first side of the disc. The sequence then repeats on the synodic period.

\begin{figure}
  \centerline{\epsfig{file=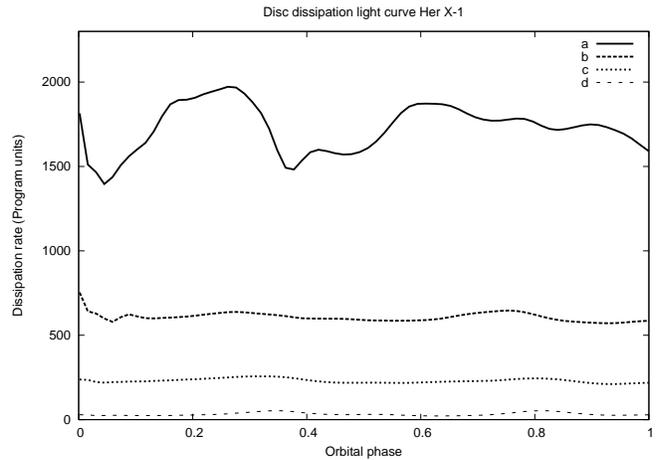,height=0.5\textwidth,angle=270}}
  \caption{
	         Her\,X-1 dissipation rate light curves for different regions of the accretion disc. Curve (a) is the dissipation rate integrated over the entire disc. Curves (b), (c) and (d) show the dissipation from the disc at distances greater $0.09a$, $0.16a$ and $0.25a$ respectively.
         }
  \label{HerX1:Figure:herx1_light_curve}
\end{figure}

The simulated disc dissipation light curves for the warped disc are shown in Figure \ref{HerX1:Figure:herx1_light_curve}. Dissipation curve (a) is the dissipation rate integrated over the entire disc and was evaluated by summing the dissipation rate from all the particles in the disc. Curves (b), (c) and (d) show the dissipation from the disc at distances greater than $0.09a$, $0.16a$ and $0.25a$ respectively. The peak at orbital phase approximately $0.3$ corresponds to when the gas stream intercepts the accretion disc close to the compact object,
labelled (a) in Figure \ref{HerX1:Figure:herx1_stream_disc};
this recurs on the $1.65$ day synodic period,
which is a few percent shorter than the orbital period.
Most of the variability in
Figure~\ref{HerX1:Figure:herx1_light_curve}
is caused by the stream impact moving close to
the compact object.
%
Figure 6 has no direct observational counterpart: for the Her\thinspace X-1 system we expect the observed optical light curve to be dominated by reprocessing of X-rays generated very close to the compact object;
modelling this reprocessing-powered optical
light curve is beyond the scope of this paper
as it would require realistic treatment of the wavelength dependent albedo.

\begin{figure}
  \centerline{\epsfig{file=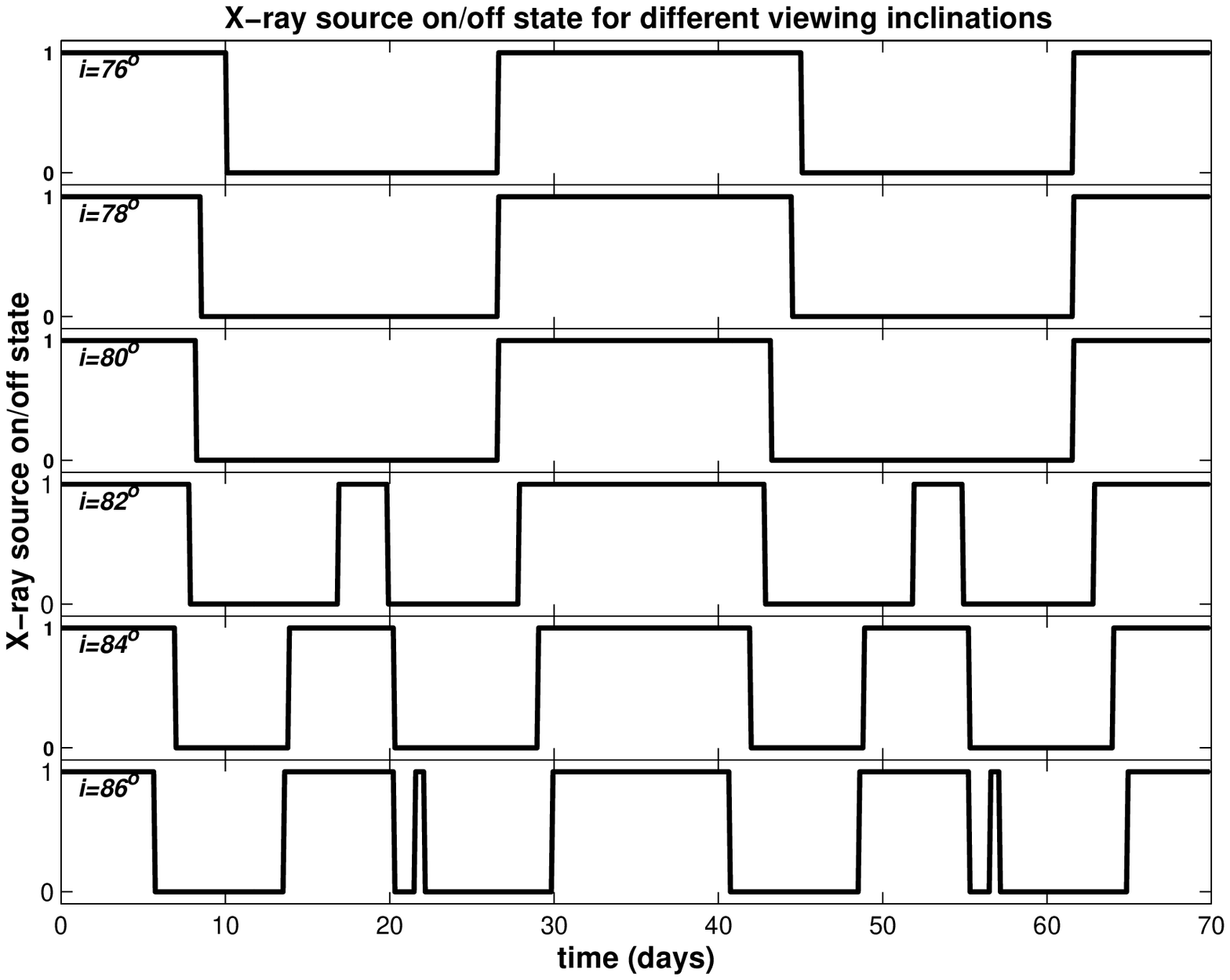,width=0.5\textwidth,angle=0}}
  \caption{
	         X-ray on and off states for two complete precession periods viewed at different inclinations for the Her\,X-1 simulation. In the graphs a value of $1$ indicates an X-ray on state, a value of $0$ is an X-ray off state. The observation angle is indicated in the upper left of each plot.
         }
  \label{HerX1:Figure:herx1_on_off}
\end{figure}

A radiation source centred on the compact object will be hidden to a high inclination observer as the tilted disc moves between the observer and the radiation source. When the line of sight is clear
the central X-ray source is visible, resulting in an X-ray on state.
However, when the disc material comes between the observer and the X-ray source the X-rays will be absorbed by the disc material leading to an X-ray off state. In Figure \ref{HerX1:Figure:herx1_on_off} we show curves marking X-ray on and off states as seen from different observer inclinations. For the first three panels
where the inclination is less than \degrees{82} only one on and off state occurs per precession period. For larger inclinations there are two on states of different lengths, one when the line of sight passes over the top of the disc and another when the line of sight passes under the disc resulting in a short on state. X-ray observations of Her\,X-1 show similar behaviour \citep{DeeterEt:1981,MiddleditchNelson:1976,GerendBoynton:1976} with a ``main-high" and a ``short-high" on states.
The inclination of $i > 82^{\circ}$ implied by our model is
consistent with independent estimates from observations of Her\,X-1.
Recent papers e.g. \citet{BorosonEt:2007} state simply $i > \degrees{80}$, and cite
\citet{HW:1983}'s value of
$i=\degrees{83.25} \pm \degrees{0.25}$ though the latter state
``we have reservations about the quoted inclination and disc tilt'' and
``it may have been preferable to adopt a fixed value for the
inclination ($i \approx \degrees{85}$)''.
Though we recognise the limitations of our model, we are encouraged that by tuning a single parameter, $L_{X}$, to reproduce the disc precession period, we simultaneously reproduce the observed X-ray luminosity, the X-ray
on/off behaviour {\bf and}
a consensus value of the orbital inclination.
We conclude that our work is strong evidence that irradiation-driven warping is the physical cause underlying the precessing tilted disc in Her\,X-1.

\subsection{SS\,433}
\begin{figure*}
  \centerline{\epsfig{file=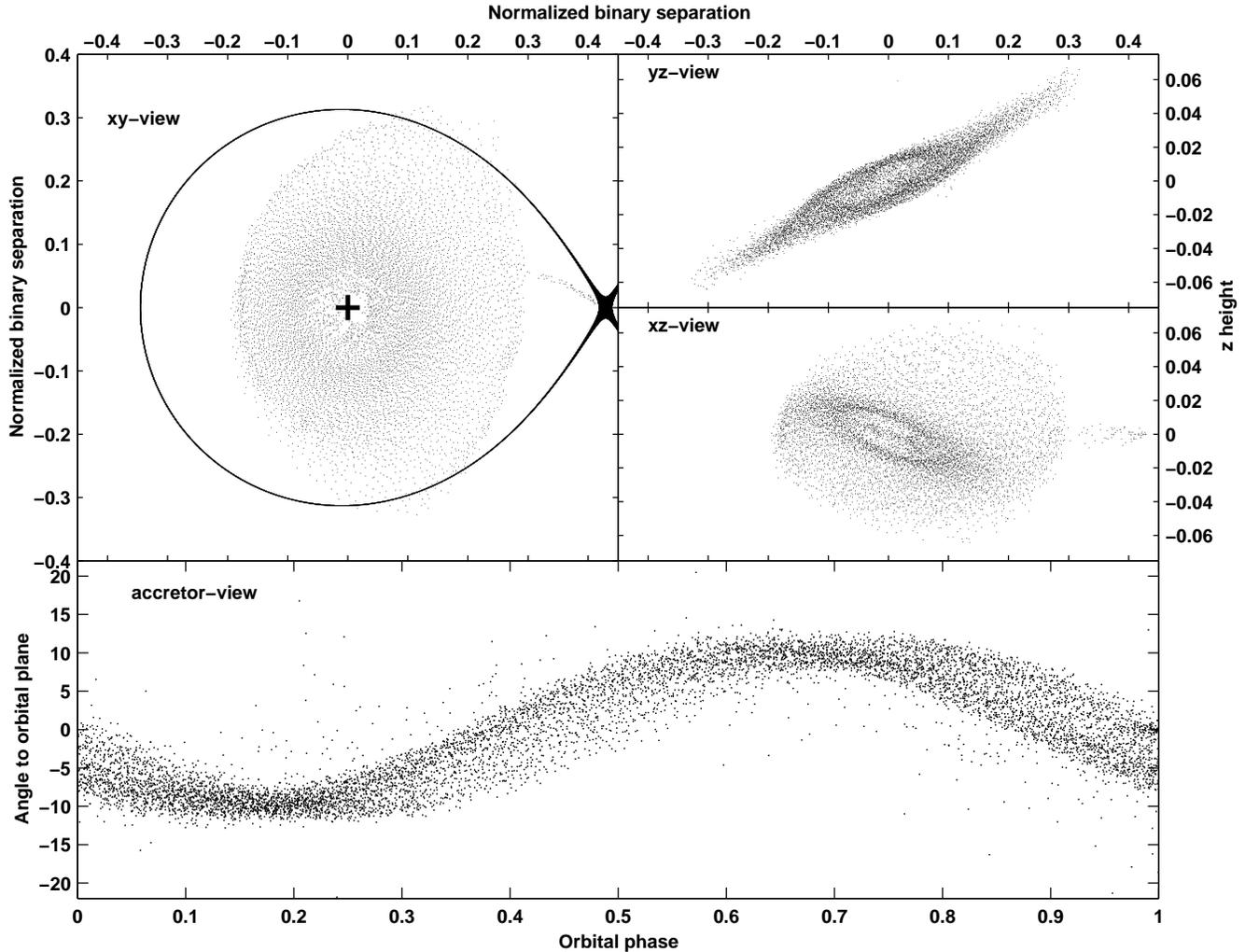,width=1.0\textwidth,angle=0}}
  \caption{
	         Particle positions plots for SS\,433. See caption of Figure \ref{HerX1:Figure:herx1_part_pos}.
         }
  \label{HerX1:Figure:ss433_part_plot}
\end{figure*}

Using parameters from Table \ref{HerX1:table:SS433Parameters}, the accretion disc for the SS\,433 model without the radiation source was also roughly point-symmetric about the compact object. The disc did not precess and remained approximately constant in shape and size once mass equilibrium had been reached. When the radiation source was switched on a strong twist developed in the disc and the entire disc tilted out of the orbital plane. Figure \ref{HerX1:Figure:ss433_part_plot} shows particle projections (as described in Figure \ref{HerX1:Figure:herx1_part_pos}) for the SS\,433 simulation. The plan view of the accretion disc shows the disc to be asymmetric. The disc had two spiral density compression waves at the edge. The plan view of the disc remained approximately constant throughout the run.
\begin{figure}
  \centerline{\epsfig{file=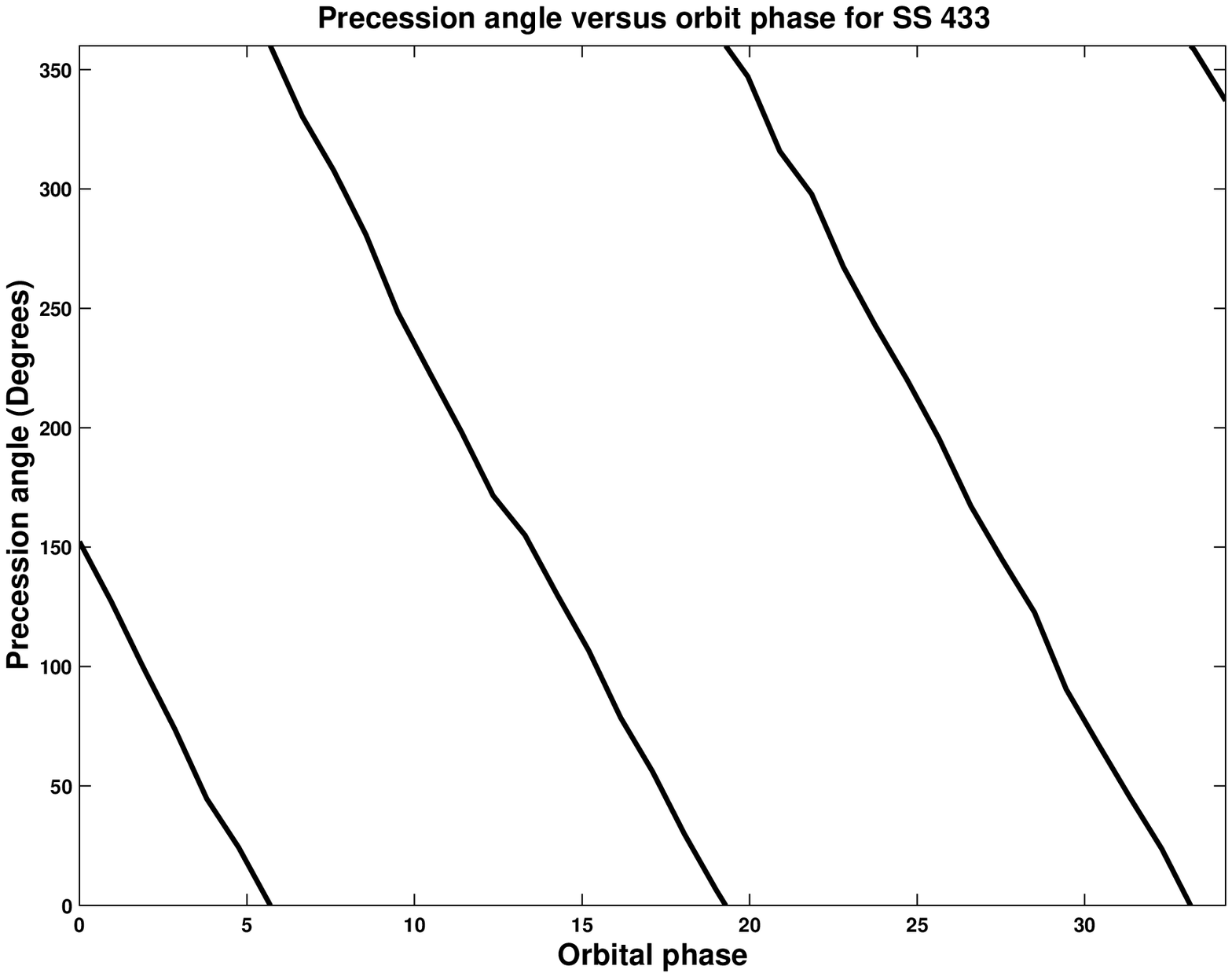,width=0.5\textwidth,angle=0}}
  \caption{
	         Disc precession angle versus orbital phase for the SS\,433 SPH simulation. The graph shows the direction of the angular momentum for the whole disc measured relative to some arbitrary starting position. The disc rotates in a retrograde direction with a precession period of approximately $162$ days.
         }
  \label{HerX1:Figure:ss433_prec_plot}
\end{figure}
The two upper right-hand plots of Figure \ref{HerX1:Figure:ss433_part_plot} show that this disc was inclined with respect to the orbital plane. In the yz-view the disc is viewed almost side on. This leads to the gas stream impacting the disc near the circularization radius instead of the edge of the disc.
The inner disc regions are relatively close to the orbital plane,
while the outer parts of the disc are tilting to large angles
\citep[c.f. Figure 7 of][]{WijersPringle:1999}.  The lower plot of Figure \ref{HerX1:Figure:ss433_part_plot}, labelled accretor-view, shows the distribution of the particles as seen from the accretor.
As in the case of Her\thinspace X-1,
the radiation force has
tilted the entire disc out of the orbital plane.
The inclination of the disc remained constant for many orbital periods and
the disc precessed in a retrograde direction, see
Figure~\ref{HerX1:Figure:ss433_prec_plot}.
We ran this simulation for 50 orbital periods in which the tilt and
warp shape remained approximately constant.
With a
central luminosity of $1.2 \times 10^{36}{\rm \, erg\,s^{-1}} $,
we obtained a
precession rate of approximately $162$ days.
This luminosity is a factor of approximately $4$ above the
2-10 keV luminosity found by
\cite{MarshallEt:2002} and about twice the 2-8 keV luminosity found by
\cite{KotaniEt:1996}: both these observed values are sensitive to the spectral model and $N_{\rm H}$. In Table~\ref{HerX1:table:SS433Parameters} we quote the
(model-dependent) luminosity from
\cite{BrinkmannEt:1991}, which is 20\% less than the $L_{\rm X}$ we obtain by tuning to the disc precession period.
More important than these discrepancies in luminosity is
the kinetic power in the jets, which is approximately $1000$ times the X-ray luminosity.
We surmise that this will have a significant effect on the dynamics of
the inner disc, but any detailed analysis requires a more definite
picture of the jet acceleration mechanism than has yet emerged.
As noted earlier in our discussion of Her X-1, our assumptions about
the albedo must also be considered when comparing our model
luminosities to the observed luminosities: since we assume zero albedo,
we are effectively boosting the irradiation by a factor of approximately $2$
at a given luminosity.
\begin{figure*}
  \centerline{\epsfig{file=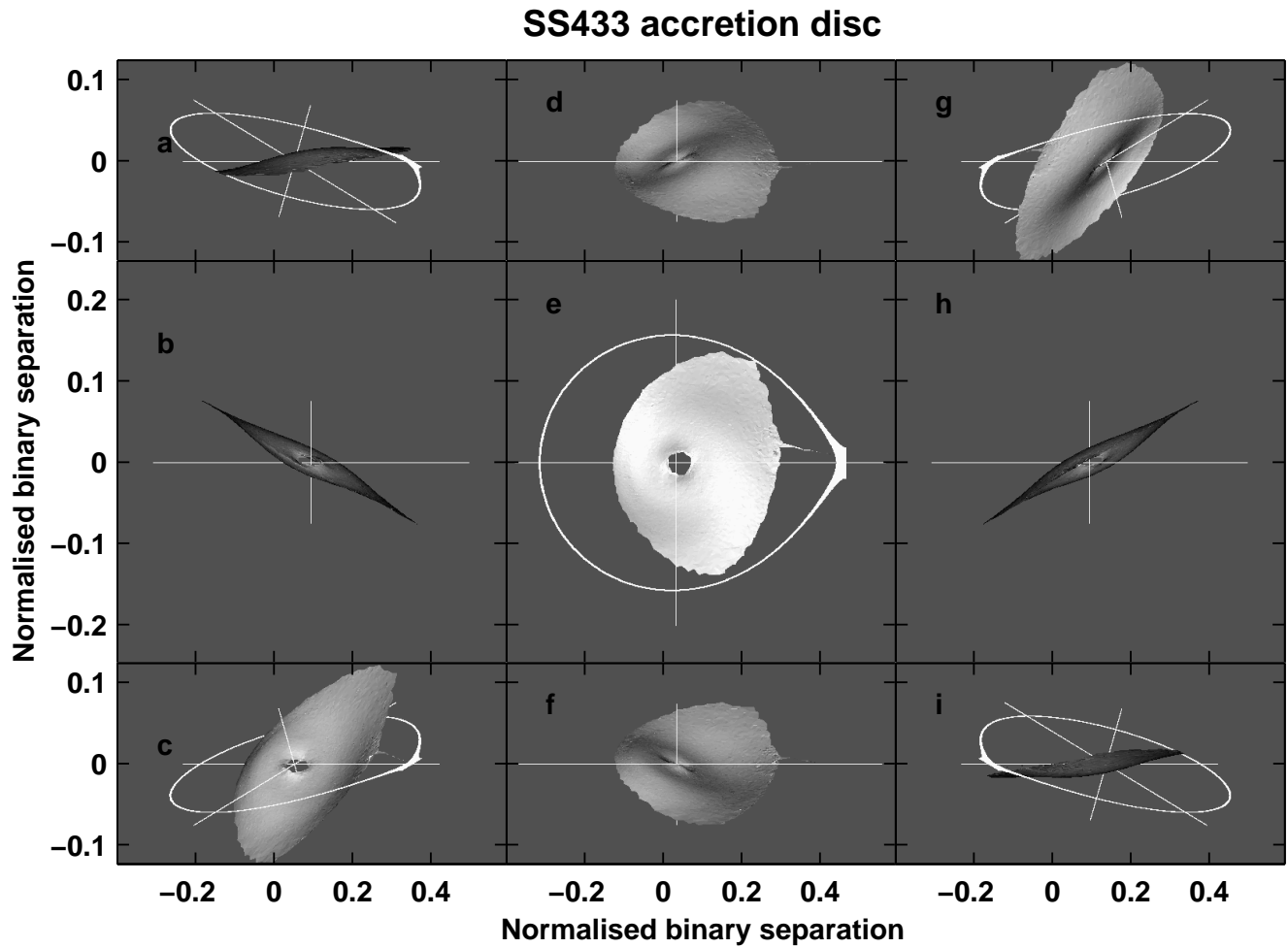,width=1.0\textwidth,angle=0}}
  \caption{
	         SS\,433 SPH accretion disc viewed at different angles,
the z values have been artificially magnified by a factor of two, so the tilt
angles here are misleading. See caption of Figure \ref{HerX1:Figure:herx1_geomview}.
         }
  \label{HerX1:Figure:ss433_geomview}
\end{figure*}


This warped disc is also inclined relative to the orbital plane as shown in
Figures~\ref{HerX1:Figure:ss433_part_plot} and \ref{HerX1:Figure:ss433_geomview}.
As illustrated for Her\thinspace X-1 in
Figure~\ref{HerX1:Figure:herx1_stream_disc},
as the secondary in SS\thinspace 433 orbits the tilted disc, the impact point of the gas stream varies from the edge of the disc to nearer its centre.

\begin{figure}
  \centerline{\epsfig{file=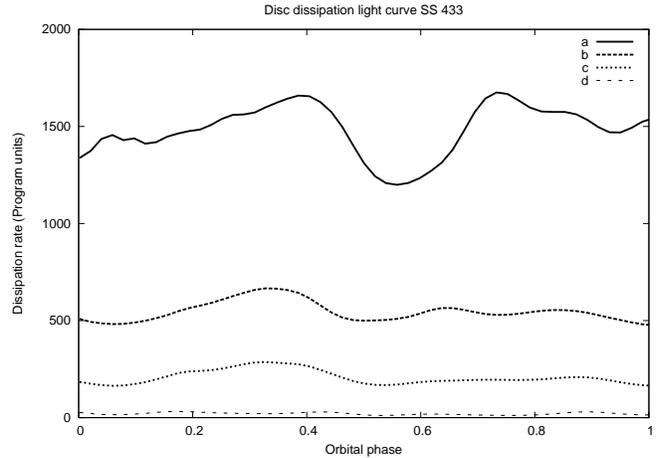,height=0.5\textwidth,angle=270}}
  \caption{
	         SS\,433 dissipation rate light curves for different regions of the accretion disc. See caption of Figure \ref{HerX1:Figure:herx1_light_curve}.
         }
  \label{HerX1:Figure:ss433_light_curve}
\end{figure}

The simulated disc dissipation light curves for the SS\thinspace 433
warped disc are shown in Figure \ref{HerX1:Figure:ss433_light_curve}. Dissipation curve (a) is the dissipation rate integrated over the entire disc. Curves (b), (c) and (d) show the dissipation from the disc at distances greater $0.09a$, $0.16a$ and $0.25a$ respectively. The dip at orbital phase approximately $0.55$ corresponds to when the gas stream intercepts the accretion disc edge. The peaks at orbital phase approximately $0.4$ and approximately $0.7$ are due to gas stream impacts
close to the central object.
The peaks and dips repeat on the
synodic period, as they did for Her\thinspace X-1, and again
most of the variability in dissipation is
generated by the stream impact moving close to the compact object.

The overall disc tilt out of the orbital plane was found to be $\degrees{11.3}$ with a maximum warp tilt angle of $\degrees{18.6}$ near the edge of the disc. The tilt of the disc will lead to the radiation source being hidden to a high inclination observer as the tilted disc moves between the observer and the radiation source. This will
occur as the warped disc precesses, as it did in the Her\,X-1 simulation
(see Figure \ref{HerX1:Figure:herx1_on_off}). The radiation source may also be obscured by the material in the warp close to the centre of the disc.

By tuning the central luminosity we have been able to reproduce the observed long period
for SS\,433. The tilt angle of the disc at the edge is very similar to the half-angle ($\degrees{20}$) for the collimated precessing jets. In the centre of the disc, which is presumably where the jets are
launched, the disc tilt angle was, however, about $\degrees{12}$.
It is intriguing to wonder whether this discrepancy might be diminished with the system parameters recently published by \cite{Blundell:2008}. We note also, however, that there is evidence for an accretion disc wind from this system, which along with the acceleration of the relativistic jets, is absent in our model.

\subsection{Other systems modelled}
\label{otherresultssec}
We also modelled eight other systems, Tables \ref{HerX1:table:OtherSystemsParameters} and \ref{HerX1:table:OtherSystemsParameters2} contain the main system parameters adopted in this study for these systems.
The eight models simulate five X-ray binaries with super-orbital periods,
the source 4U\,1626-67 discussed by
\cite{WijersPringle:1999},
a `generic' LMXB, and the CV KR\,Aur.
For the `generic' LMXB, $P_{\rm long}$ is that of our model.
KR\,Aur is a `negative superhump'
system which
\cite{Kozhevnikov:2007} suggests may harbour a radiation-driven warped disc.
In each relevant case,
the model luminosities
were chosen to reproduce the observed long periods found in these systems.
In all cases the disc developed strong warps,
 detailed in Table \ref{HerX1:table:OtherResults}.

\begin{table*}
\caption
  {
    Major parameters for the other binary systems modelled
  }
\label{HerX1:table:OtherSystemsParameters}
\begin{center}
\begin{tabular}{c c c c c c c}
\hline
System     & $P_{orb}$  & q             & $M_1$     & $\dot{M}_{s}$               & $P_{\rm long}$ & Reference \\
name       & (days)     & $M_{2}/M_{1}$ &$M_{\odot}$&($10^{-8}M_{\odot}\,{\rm yr^{-1}}$) & (days)     &           \\
\hline
\hline
SMC X-1    & 3.9        & 12.9          & 1.4       & 4.0                         & $\approx55$   & 1         \\
Cyg X-1    & 5.6        & 2.1           & 16.0      & 2.5                         & 294        & 1         \\
Cyg X-2    & 9.844      & 0.5           & 1.4       & 1.0                         & 78         & 1         \\
X 1916-053 & 0.035      & 0.07          & 1.4       & 0.1                         & 3.8        & 1         \\
4U 1626-67 & 0.029      & 0.02          & 1.4       & 0.05                        & 0.0        & 1         \\
LMC X-3    & 1.7        & 0.6           & 10.0      & 3.0                         & 198        & 1         \\
KR Aur     & 0.16       & 0.6857        & 0.7       & $1.0\times 10^{-10}$  			 & 5.3	    	 & 2         \\
Gen LMXB   & 0.2        & 0.36          & 1.4       & 0.1                         & 5.25       & 1         \\
\hline
\end{tabular}
\end{center}
Reference: $^1$ \cite{WijersPringle:1999}, $^2$ \cite{Kozhevnikov:2007}
\end{table*}

\begin{table*}
\caption
  {
    X-ray Luminosity and system type
  }
\label{HerX1:table:OtherSystemsParameters2}
\begin{center}
\begin{tabular}{c c c c c c}
\hline
System     & X-ray Luminosity  $L_*$         & $L_{Edd}$                         & Type & Reference \\
name       & ($\times10^{38}$ $erg\,s^{-1}$) & ($\times\,10^{38}$ $erg\,s^{-1}$) &      &           \\
\hline
\hline
SMC X-1    & 2.0                             & 1.82                              & HMXB & 1         \\
Cyg X-1    & 0.05                            & 20.8                              & HMXB & 1         \\
Cyg X-2    & 1.8                             & 1.82                              & LMXB & 1         \\
X 1916-053 & 0.05                            & 1.82                              & LMXB & 2         \\
4U 1626-67 & 0.1                             & 1.82                              & LMXB & 3         \\
LMC X-3    & 1.5                             & 13.0                              & HMXB & 1         \\
KR Aur     & see text                              &    see text                             & LMXB & 4         \\
Gen LMXB   & 1.0                             & 1.82                              & LMXB & 1         \\
\hline
\end{tabular}
\end{center}
Reference: $^1$ \cite{GrimmEt:2002}, $^2$ \cite{NaritaEt:2003}, $^3$ \cite{Chakrabarty:1998}, $^4$ \cite{Kozhevnikov:2007}
\end{table*}

For systems with extreme mass ratios, i.e. where $M_{\rm compact} >$ approximately $4 M_{\rm donor}$
\citep[c.f.][]{Smithench:2007}
a strong warp develops close to the centre of the disc to
the extent that some of the outer parts
are completely shadowed by the inner warp. The warp precesses in a wave-like manner in a retrograde
direction relative to the inertial frame.
In addition,
the outer disc precesses in a prograde direction due to the well-known
eccentric instability first discovered in the SU\thinspace UMa CVs
\citep[see e.g.][]{Vogt:1982,Smithench:2007}.
In our extreme mass-ratio warped disc simulations (i.e. those
for X\thinspace 1916-053 and
4U\thinspace 1626-67) superhumps are apparent in the simulated
dissipation light curve, just as they were for the
plane disc simulated by \citet{FoulkesEt:2004}.
We found a similar behaviour in the simulations of Bo\thinspace 158
\citep{BarnardEt:2006}.
For the less extreme mass ratio systems a small warp is induced in the inner regions of the disc. This warp does not shadow the outer regions of the disc and these regions also warp. The entire disc then becomes tilted to the binary plane and the whole disc precesses in a retrograde direction.
This is reflected in Table~\ref{HerX1:table:OtherResults}: the two extreme mass ratio systems have their maximum warps in the inner disc. Generally the other systems have their maximum warp at the disc edge, but the Cyg\,X-1 model is an exception to this rule.

4U\thinspace 1626-67 was modelled to examine the hypothesis of
\cite{KerkwijkEt:1998}, \cite{WijersPringle:1999} and others
that for large central luminosities the inner disc warp angle exceeds
$\degrees{90}$, so these disc annuli flow in a direction counter to the orbital angular velocity. If this were the case, the reversal of the disc angular
momentum
vector could account for the torque reversals observed
by e.g. \cite{ChakEt:1997}. As Table~\ref{HerX1:table:OtherResults} shows,
we find an inner disc warp of $\degrees{21}$ for 4U\thinspace 1626-67.
This is the largest angular deviation from the orbital plane found in
any of our simulations, so we agree with \cite{WijersPringle:1999} that
4U\thinspace 1626-67 is an extreme case. We suggest, however, that the
inversion of the disc angular momentum vector in their inner disc is simply
an artefact of the approximations necessitated by their analytic approach.
Our conclusion leaves the mechanism underlying the observed torque reversals
as an open question.

The luminosity used in each system model was fine-tuned
to reproduce the observed long periods
and tabulated in Table~\ref{HerX1:table:OtherResults}.
For all eight X-ray binaries with observed super-orbital periods, the luminosities required to produced the observed long periods
are similar to the observed luminosity\footnote{In the tables and Fig.~\ref{HerX1:Figure:ll_log} we have used the long-term average luminosity, derived from the RXTE ASM, where this has been published. For three of the four brightest sources the error on this value is smaller than the size of the symbol used. For the fourth, we used the data of Staubert et al (2007) to deduce the size of the error bar, again finding it to be smaller than the symbol used. For the three least luminous sources we were unable to a reliable error bar for the long-term average luminosity.
 }, see Figure~\ref{HerX1:Figure:ll_log}.
The biggest discrepancies are about 20\% in the cases of SS\,433 and SMC\,X-1.
As already noted, our model is of limited applicability to SS\,433 because we do not attempt to model the kinematic luminosity. SMC\,X-1 is probably partly powered by wind-fed accretion, altering the angular momentum balance in the disc from that in our exclusively Roche lobe overflow fed model disc.
Given the simplistic treatment of the reprocessing and the limitations in our simulation of the accretion disc viscosity, this agreement is astonishingly
good. We conclude that irradiation driven warping is probably generally
the mechanism underlying the observed long periods in X-ray binaries.

For KR\,Aur the X-ray luminosity is poorly known. \cite{Shafter:1983}
cites a private communication giving the Einstein
IPC 0.15 -4.5 keV count rate as
$7.9 \times 10^{-2}{\rm \, counts \, sec^{-1}}$.
This implies a $L_{X}= 10^{30.7} {\rm \, erg \, sec^{-1}}$
using a distance of 180\,pc \citep{Greiner:1998}.
The ASCA 0.7 - 10keV blackbody fit (a poor fit to the spectrum)
gives $L_{X} = 10^{30.4} {\rm \, erg \, sec^{-1}}$
using a distance of 100 pc or $L_{X} = 10^{30.9} {\rm \, erg \, sec^{-1}}$ using a distance
of 180 pc.
These numbers are very low compared to the luminosities mentioned in the final discussion
in \cite{Kozhevnikov:2007}, and are indeed typical of the
X-ray/optical flux ratios of optically selected
CVs \citep{Shafter:1983, Patterson:1981}.

As Table~\ref{HerX1:table:OtherResults} shows, the model $L_{X}$ required to match the inferred warp disc precession period in KR\,Aur is $10^{37}{\rm \, erg \, sec^{-1}}$. Intriguingly, this is typical of the X-ray luminosity for supersoft sources \citep{Greiner:2000}.

\begin{table*}
\caption
  {
    Results the other binary systems modelled. Columns are: 1 System modelled, 2 Model luminosity used to reproduce the system long period, 3 Maximum disc warp angle, 4 Position of the maximum disc warp, Inner disc or Edge of the disc and 5 The overall disc tilt.
  }
\label{HerX1:table:OtherResults}
\begin{center}
\begin{tabular}{ c c c c c}
\hline
System     &  Model Luminosity $L_*$            & Maximum Disc Warp  & Maximum Disc  & Disc tilt \\
name       &  ($\times\,10^{38}$ $erg\,s^{-1}$) & (Degrees)          & Warp Position & (Degrees) \\
\hline
\hline
SMC X-1    & 1.65                               & 20.9               & Edge          & 12.3      \\
Cyg X-1    & 0.05                               & 14.7               & Inner         & 6.2       \\
Cyg X-2    & 1.70                               & 19.4               & Edge          & 10.5      \\
X 1916-053 & 0.05                               & 12.8               & Inner         & 0.8       \\
4U 1626-67 & 0.1                                & 21.0               & Inner         & 1.0       \\
LMC X-3    & 1.75                               & 19.6               & Edge          & 10.6      \\
KR Aur     & 0.1                                &  7.5               & Edge          & 3.0       \\
Gen LMXB   & 1.0                                & 18.5               & Edge          & 10.6      \\
Her\,X-1   & 0.215                              & 13.7               & Edge          &  5.5      \\
SS\,433    & 0.012                              & 18.6               & Edge          & 11.3      \\
\hline
\end{tabular}
\end{center}
\end{table*}

\subsection{Orbital inclinations and X-ray Dips}
\label{dipsec}
In LMXBs X-ray dips are well-understood phenomena, arising when material from the accretion flow transiently blocks our line of sight to the
central X-ray source.
Table~\ref{HerX1:table:OtherResults}
gives the maximum warp angle and overall disc tilt angle
for each system simulated. These results allows us to place constraints on the systems' orbital inclinations
using observations of the presence or absence of X-ray dips. Or, alternatively we can use the presence or absence of X-ray dips, coupled with known orbital inclinations, to examine whether our results make sense.
Her\,X-1 and SMC\,X-1 are eclipsing, so the inclinations are strongly constrained by this. We have already shown that the disc shape we find for Her\,X-1 is consistent with the X-ray on and off behaviour for $i \approx \degrees{82}$.

Of the LMXBs we treat, all show X-ray dips.
Since we selected them for their observed super-orbital periods
and well-constrained system parameters
(except for  4U\,1626-67, selected because
previous authors discussed it in the context
of warped precessing discs)
this is interesting. It implies that the long period modulations may be
preferentially  detected at high orbital inclinations.
This is consistent with the moderate values we find for
the angular deviation of the disc from the orbital plane.
The biggest
overall disc
tilt is $\degrees{12.3}$ for SMC\,X-1
while the most pronounced angular deviation is the
$\degrees{21}$ inner warp in 4U\,1626-67
(see Table~\ref{HerX1:table:OtherResults}).
At low inclinations the disc will always appear
more-or-less the same if the overall disc tilt has the moderate values we find.
In contrast, some of the models discussed in the literature for Her\,X-1 have tilts of $\degrees{40}$: if this were typical, then the precession would produce
observable effects at low and moderate inclinations.
As a caveat we note that system parameters are generally easier to determine
in high inclination systems, so the preponderance of dippers may be partly due
to this selection criterion.

4U\,1916-053 is an extensively-studied dip source, without a well-known orbital inclination. Our results, with a maximum disc warp of $\degrees{12.8}$,
suggest an orbital inclination of approximately $\degrees{77}$, such that the inner disc warp intermittently
blocks the line of sight to the central X-ray source.
4U\,1626-67 is similarly a well-known dip source. In this case, we find a pronounced inner disc warp of $\degrees{21}$, the highest for all our simulations. This implies that the orbital inclination in this case is less than approximately $ \degrees{70}$.

Cyg\,X-2, conversely, is not a well-known X-ray dip source. Dips were reported by \citet{VrtilekEt:1986}, but despite Cyg\,X-2 being one of the brightest persistent LMXBs we did not find other dips in the literature. This suggests
Cyg\,X-2 has an inclination which is just lower than that required for dipping
to be seen, and only when stochastic variations happen to conspire to move
material an unusual distance above the disc mid-plane do dips occur.
The inclination of Cyg\,X-2 was studied by \cite{OroszKuulkers:1999}.
They find  $i \approx \degrees{55}$, from modelling of the
ellipsoidal variations assuming the disc has a temperature profile
$T \propto r^{-{3/7}}$
typical of steady irradiated discs; or $i=\degrees{62.5} \pm \degrees{4}$ assuming the canonical steady-state disc $T \propto
r^{-{3/4}}$.
Coupled with our maximum disc tilt of $\degrees{19.4}$, these numbers are consistent with dipping only in rare, extremely favourable stochastic deviations.

Taken overall, therefore, we consider our results for the LMXBs to be encouragingly
consistent with the observational literature. Emboldened by this, we take
our values for the maximum disc warps in
4U\,1916-053 and 4U\,1626-67 at face value, and venture to predict that the inclinations in these two systems are approximately $\degrees{77}$ and approximately $\degrees{69}$. These are simple estimates, ignoring the effect of the finite opening angle of the disc. If the warped disk maintains the canonical $H/R \approx 0.1$ relationship, then our inclination estimates would be reduced by approximately $\degrees{6}$ in each case.

We also examined the literature for dips in the HMXBs, finding nothing except in the case of Cyg\,X-1. In this case the orbital inclination is accepted to be approximately $\degrees{30}$ (e.g. \citep{GiesBolton:1986}) and the dips are likely associated with material in the wind from the donor star.
Where discs are wind-fed, the specific angular momentum and the X-ray luminosity is generally less.
Thus HMXB discs are generally smaller and less irradiated than LMXB discs, both characteristics tending to reduce the angular elevation of outlying material above the disc mid-plane. Our simulations show that for HMXB parameters, the discs invariably have their maximum warps at the outer edge, lacking the pronounced inner disc warps of the extreme mass ratio LMXBs. All of these factors conspire to make dips from accretion disc material in HMXBs less prevalent than in LMXBs.

\section{Conclusions}
\label{HerX1:sec:Conclusions}

We have studied the radiation-driven warping of thin accretion discs in
X-ray binaries  and one CV
using a non-linear SPH code.
A convex hull algorithm was used to find the surface particles of the disc and a ray-tracing algorithm used to evaluate regions of self-shadowing, as described fully in Paper I. Initially planar accretion discs were illuminated with radiation from the centre of the discs. The surface of each disc was subject to a force from the absorption of the X-rays on the disc surface and back reaction from the remitted radiation.

The radiation caused these discs to warp and
tilt out of the orbital plane, precessing
in a retrograde direction relative to the inertial frame. The resulting tilt and precession rate remained constant for many orbital periods.

\begin{figure}
  \centerline{\epsfig{file=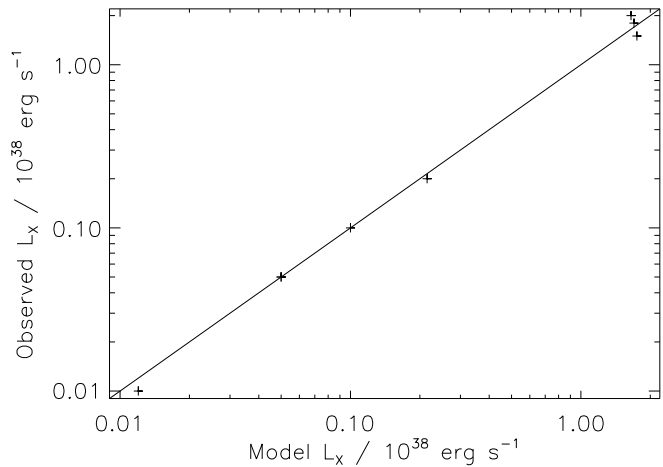,height=0.39\textwidth,angle=0}}
\caption{
                 The observed X-ray luminosity
versus the model X-ray luminosity required to match the observed
super-orbital period for the seven systems with observed super-orbital periods
(crosses). The line is Model $L_{\rm X} = $ Observed  $L_{\rm X}$.
In all cases the model
X-ray luminosity required to
match the observed super-orbital period is extremely close to the observed X-ray luminosity. We consider this excellent agreement to be strong evidence in favour of the hypothesis that these super-orbital periods are the result of precession of irradiation-driven warped discs.}
  \label{HerX1:Figure:ll_log}
\end{figure}

We simulated a number of systems that have long
periods in their observed light curves. We
tuned the irradiating luminosity, $L_{*}$,
to match the precession-rate of the warped disc to the observed long periods in seven X-ray binaries and the inferred warped disc precession period in
one CV.
Figure~\ref{HerX1:Figure:ll_log}
shows the astonishingly good match between observed $L_{\rm X}$ and the irradiating luminosity, $L_{*}$, demanded by our models to match the observed periods.
The largest deviations from the 1:1 straight line in
Figure~\ref{HerX1:Figure:ll_log}
are approximately $20\%$, less than the typical uncertainty introduced by the spectral model and the distance determination.

\cite{OgilvieDubus:2001} [OD01] found that only low-mass X-ray binaries with high orbital periods are likely to be unstable to irradiation-driven warping. In contrast to this, our simulations find irradiation-driven warping for all the orbital periods considered. In common with us, OD01 considered self-shadowing and non-linear solutions. Their work is analytical and consequently necessarily makes approximations. Our simulations are able to capture more complex behaviour. For example OD01 state ``Owing to the symmetries of the problem, these non-linear solutions are steady in the rotating frame of reference..." whereas our disc continuously flexes in response to the changing orientation of the Roche potential. Our short orbital period LMXBs tend to have extreme mass ratios, and for these the motion is particularly complex, with prograde apsidal precession in the outer disc, warped retrograde precession in the inner disc and the spiral structure all contributing to a changing self-shadowing of the disc surface. Using analytical techniques it is unfeasible to robustly capture these complex motions on a variety of timescales.

\begin{figure}
  \centerline{\epsfig{file=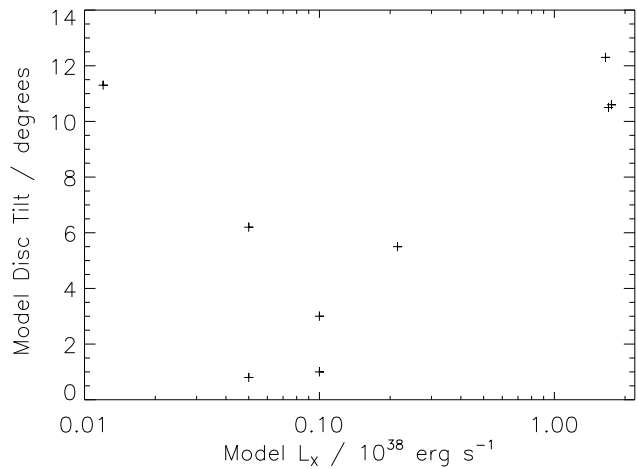,height=0.39\textwidth,angle=0}}
\caption{
                 The model overall disc tilt versus the model X-ray
luminosity. The scatter of points indicates that parameters other than $L_{X}$ play an important role in determining the disc tilt.
}
  \label{HerX1:Figure:ltilt_log}
\end{figure}

\begin{figure}
  \centerline{\epsfig{file=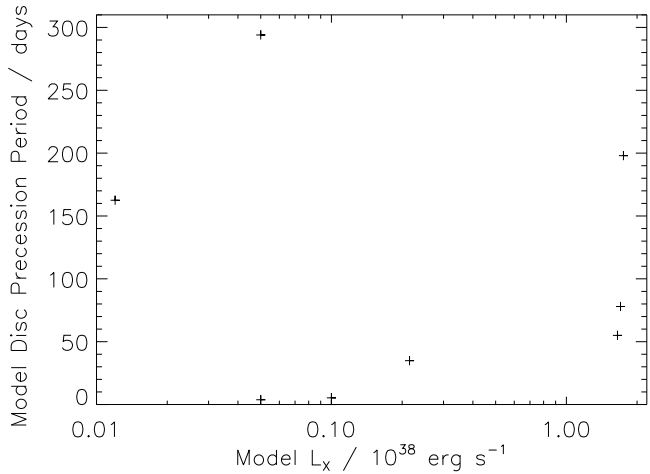,height=0.39\textwidth,angle=0}}
\caption{
                 The model disc precession period  versus the model X-ray
luminosity. The scatter of points indicates that parameters other than $L_{X}$ play an important role in determining the disc precession period.
}
  \label{HerX1:Figure:lplong_log}
\end{figure}

In Figures~\ref{HerX1:Figure:ltilt_log} and ~\ref{HerX1:Figure:lplong_log}
we show how the disc tilt and the disc precession periods depend on the
model luminosity, $L_{*}$. The scatter in these two plots demonstrate that
the disc behaviour is influenced by parameters other than luminosity in a complex way.
This strengthens our confidence in our conclusion that
Figure~\ref{HerX1:Figure:ll_log} provides strong evidence in favour of the hypothesis
that the observed long periods are caused by the precession of irradiation-warped discs.

Her\,X-1 is probably the best-characterised of the observed systems.
Our simulation of its disc appears to simultaneously
match several of the system characteristics by tuning a single parameter,
the X-ray luminosity.
Encouragingly, the tuned value for this parameter is consistent with the observed value. Our Her\,X-1 disc produces the observed ``main-high'', ``short-high'' and off states for an orbital inclination consistent with independent determinations.
As recently as 2007, \citet{BorosonEt:2007} stated that the physical cause for the warped precessing disc remains unknown. We strongly conclude that irradiation-driven warping underlies the long-known and well-observed tilted precessing disc in Her\,X-1.

For the other X-ray binaries, many of the parameters are subject to significant uncertainty. Encouraged by our  Her\,X-1 results, we surmise that we may be able to use the results in Table 5 to constrain the system orbital inclinations, using the observed presence, absence and periodicities in X-ray dipping behaviour.
We discussed this in section~\ref{dipsec} finding our results
to be consistent with the observations for all the LMXBs. We venture to predict
inclinations for X\,1916-053 and 4U\,1626-67 of approximately $\degrees{77}$ and $\degrees{69}$ respectively. We also noted that we find maximum warps located in the inner discs for systems like these two, which have mass ratio extreme enough for the disc to harbour the 3:1 resonance, leading to an apsidally precessing eccentric outer disc. The general lack of these pronounced inner disc warps for less extreme mass ratios may partially explain the lack of X-ray dips in HMXB systems, as discussed at the end of section~\ref{dipsec}.

We included a simulation of the CV KR\,Aur, motivated by the statements in
\cite{Kozhevnikov:2007}. We found we could reproduce the inferred tilted disc precession period for this system, but only by using $L_{X} =10^{37}
{\rm \, erg \, sec^{-1}}$. This value is about a million times higher than the observed values of $L_{X}$ we were able to find for this system,
but is consistent with the X-ray luminosities of supersoft sources. Thus
our findings seem consistent with the hypothesis advanced by \cite{Kozhevnikov:2007} that disc precession could be present in
KR\,Aur and other VY Scl stars if they
harbour white dwarfs with steady nuclear burning on their surfaces.
The luminosity thus generated
is sufficient to produce irradiation-driven warping and tilting of
the accretion disc which precesses and leads to the observed negative superhumps. The obvious test of this hypothesis is to verify that the inferred
nuclear-burning luminosity does indeed exist.

For Her\,X-1
we showed that
the position of the stream impact on
the warped and tilted disc was a function of both orbital and
disc precession phase.
Generally the  gas stream misses the edge of the disc, which is tilted out of the orbital plane, as illustrated in Figures 1 and 2.
The stream flows deeper into the gravitational well of the primary,
and hits the face of the disc at a continually-varying distance
from the compact object, as illustrated in Figure 5.
We showed the resulting
simulated disc dissipation light curve in Figure 6, but we note that for
Her\,X-1 itself, this curve is not directly comparable to any observable.
This is because Her\,X-1 is a relatively large binary, and the observed optical light will be dominated by reprocessing of the X-rays generated
well within the closest approach of the stream-disc impact point.

There is however, an interesting possible application of the varying stream
impact depth: this impact may power X-rays directly in some very compact
X-ray binary systems. This would require disc warping to allow the stream to penetrate close to the compact object, and is therefore similar in flavour to the direct-impact accretor model discussed by \cite{MarshEt:2004}.

\section{Acknowledgments} SBF acknowledges the support from QinetiQ, Malvern. We acknowledge support from the Open University's Research School and The Swinburne Centre for Astrophysics and Supercomputing facility.  We thank Will Clarkson and Jim Pringle for comments.

\label{lastpage}

\end{document}